\documentclass[amsmath,amssymb,floatfix,aps,nofootinbib,notitlepage,superscriptaddress,twocolumn,pra]{revtex4-1}
\usepackage{amsmath,amsthm,amssymb,bm,hyperref,graphicx,color,float}

\allowdisplaybreaks

\newtheorem{thm}{Theorem}
\newtheorem{lem}{Lemma}

\newcommand{\be}{\begin{equation}}
\newcommand{\ee}{\end{equation}}
\newcommand{\bea}{\begin{eqnarray}}
\newcommand{\eea}{\end{eqnarray}}
\newcommand{\6}{\partial}

\newcommand{\inti}{\int_{-\infty}^{+\infty}}

\begin{document}

\title{Non-equilibrium dynamics of the anyonic Tonks-Girardeau gas at finite temperature}

\author{Ovidiu I. P\^{a}\c{t}u}
\affiliation{Institute for Space Sciences, Bucharest-M\u{a}gurele, R 077125, Romania}

\begin{abstract}

We derive  an exact description of the non-equilibrium dynamics at finite temperature for the anyonic Tonks-Girardeau
gas extending the results of  Atas \textit{et al.} [Phys. Rev. A {\bf 95}, 043622 (2017)] to the case of arbitrary
statistics. The one-particle reduced density matrix is expressed as the Fredholm minor of an integral operator with
the kernel being the one-particle Green's function of free fermions at finite temperature and the statistics parameter
determining the constant in front of the integral operator.
We show that the numerical evaluation of this representation using Nystr\"{o}m's method significantly outperforms the
other approaches present in the literature when there are no analytical expressions for the overlaps of the wave-functions.
We illustrate the distinctive features and novel phenomena present in the dynamics of anyonic systems in two experimentally
relevant scenarios: the quantum Newton's cradle  setting and the breathing oscillations initiated by a sudden change of
the trap frequency.

\end{abstract}

\maketitle

\section{Introduction}

In recent years considerable efforts have been made in order to understand the principles underlying
the non-equilibrium dynamics of integrable and non-integrable isolated many-body systems. The results of a
large body of literature can be succinctly summarized as follows. After a quench non-integrable (chaotic)
systems are described in the long-time limit by a Gibbs (thermal) ensemble \cite{Deutsch,Sred,RDO08}.
Integrable systems, however fail to thermalize and their asymptotic properties are described by a
Generalized Gibbs Ensemble (GGE) \cite{RDYO,VR16} which needs to take into account all the local and
quasi-local integrals of motion of the system \cite{INWCE,IQNB,IMPZ}. In the case of near-integrable
systems we encounter the phenomenon of prethermalization  \cite{MMGS,EKM,BCK15,BF15,BEGR15,MMAS16,LGS16,AF17}
which is characterized by a quasi-stationary state with almost integrable features at short to moderate times
reaching thermal equilibrium only after very long time.

The main  impetus behind these theoretical developments was a series of experiments with ultracold atomic
gases  \cite{HLFSS,TCFMS,GKLKS,LGKRS,LEGRS,NKHJM,CBPES,RSBHL,FCJB} of which the quantum Newton's cradle (QNC)
experiment of  Kinoshita \textit{et al.} \cite{KWW07} stands out.
Trapped ultracold atomic gases represent the perfect testing ground for the investigation of non-equilibrium
dynamics of quantum many body systems due to their unprecedented level of control of control over interactions,
dimensionality and even statistics which allows for the realization of integrable and non-integrable systems
which can be accurately monitored over long time scales. In addition these systems are also characterized by
weak coupling to their environment which means that to a good approximation they can be considered as isolated.

One of the most important models that can be realized with ultracold atoms is the Lieb-Liniger model \cite{LL63}
which describes one-dimensional bosons with repulsive contact interactions. In a homogeneous system  and for
arbitrary values of the repulsion the Lieb-Liniger model is integrable and the wave-functions, energy spectrum
and low-lying excitations can be obtained using the Bethe ansatz \cite{LL63,L63,KBI}. In typical experiments
the system is confined to one-dimension by using a strong transverse optical trap while in the longitudinal
direction there is a harmonic potential which breaks integrability. In the limit of infinite repulsion between
the particles the Lieb-Liniger model describes the so-called Tonks-Girardeau gas \cite{Gir60} which is integrable
in both  homogeneous and inhomogeneous case using the Bose-Fermi mapping \cite{Gir60,RCB,GWT01,GW00a,GW00,DGW02,YG05}.

A natural generalization of the Lieb-Liniger model in the case of arbitrary statistics is given by the anyonic
Lieb-Liniger model introduced in \cite{Kundu} (see also \cite{BGO,BGH}). This model has been studied intensely in
the last decade and a large body of knowledge has been steadily accumulating including the properties of the
ground-state \cite{BGO,HZC1}, form factors \cite{PSC00}, the asymptotic behavior of the correlation functions for
homogeneous \cite{CM,PKA,SC,CS1,PKA4,PKA5} and trapped systems \cite{MPC,HS,SPC20b},  and entanglement \cite{SFC,GHC}.
The non-equilibrium properties after particular quenches at zero temperature were studied in \cite{delC,WRDK,PC}.
Experimental proposals of realizing 1D anyonic
systems with ultracold atoms in optical lattices using various methods such as Raman-assisted tunneling \cite{KLMR,GS},
periodically driven lattices  \cite{SSE} or multicolor lattice-depth modulation \cite{CGS1,GCS2} have reignited
 interest in the study of systems with fractional statistics \cite{IT1,IT2,AN,Girar,PKA1,LMP,BGK,BFGLZ,BCM,Greit,Kundu2,
CMT,Belazz,BS,RFB,WWZ,MS,LH,DKPBJ,RCSM,HK,YP,YCLF,SPK,Patu1,Patu2,Zinn,GLZ20,MS20,HZC2,HC1,Hao1,ZGFSZ,AFFSV1,AFFSV2}.

The majority of the analytical and numerical investigations of the non-equilibrium dynamics in the literature
treat the case of zero temperature. However, experiments are realized at nonzero temperatures and therefore
it is important to have analytical tools to study finite temperature dynamics. Another reason why it is important
to derive finite temperature results is to establish a priori bounds on the physical parameters
for which phenomena predicted at zero or low temperatures can still be seen and not be washed out by thermal
fluctuations.

In this paper we  investigate the finite temperature dynamics of the anyonic Tonks-Girardeau gas focusing on two
experimentally relevant scenarios: the quantum Newton's cradle setting and the breathing-mode oscillations after a quench
of the trap frequency. At first glance it would seem that the knowledge of the wave-functions, which
can be obtained using the Anyon-Fermi mapping \cite{Girar,PKA1}, would make the task of computing physical
relevant observables like the momentum distribution an easy one. This is however an unwarranted assumption due
to the fact that a brute force approach of computing the one-particle reduced density matrix (RDM) requires
a computational cost which increases exponentially with the number of particles.  A numerical efficient method
to compute the dynamical  RDM  of impenetrable bosons at zero temperature was derived by Pezer and Buljan in
\cite{PB07}. The main computational cost of this method is represented by the calculation of the determinant
and inverse of a matrix whose dimension is equal to the number of particles and the elements are given by the overlaps
of the wave-functions. This method is extremely efficient when the overlaps can be computed analytically and
it was extended in the anyonic case by del Campo in \cite{delC}.

The generalization for impenetrable bosons at finite temperature was introduced rather recently by Atas \textit{et al.}
in \cite{AGBKa} (for the so-called "emergent eigenstate solution" of the bosonic TG gas valid  at any temperature see \cite{VXR17}).
They considered that in the initial state the system is described by a grand-canonical ensemble
and made use of Lenard's formula \cite{Len66} which expresses the  RDMs of bosons in terms of an infinite series
involving the  RDMs of free fermions. This series in the case of the one-particle RDM is the first Fredholm minor
of an integral operator with the kernel given by the one-particle RDM of free fermions at finite temperature. Truncating the
expression for the free fermionic RDM after a finite number of terms (this is perfectly justified because each
term is multiplied by the Fermi-Dirac occupation  factor) the authors of \cite{AGBKa} were able to obtain an
expression for the bosonic RDM which is almost identical with that derived at zero temperature in \cite{PB07}
with two differences: the wave-functions overlaps are now ``dressed" with the square root of the Fermi-Dirac
occupation factors and the dimension of the matrix now is equal with the truncation level. Our derivation
presented in this paper in the case of the anyonic Tonks-Girardeau gas follows along similar lines and makes
use of the anyonic generalization of Lenard's formula \cite{PKA1} but for the numerical treatment we prefer a
different approach.
When the overlaps can be calculated analytically  the numerical method derived in \cite{AGBKa} is almost
always preferable, however, when this is not possible, which is the case in many physical relevant situations
like in the QNC setting, a more efficient numerical method is based on the evaluation of the
Fredholm minor using Nystr\"{o}m's method \cite{PFTV,Born}. A detailed comparison of the two methods can be
found in Appendix \ref{a4}  where it is shown that for moderate or large number of particles and especially at
finite temperatures Nystr\"{o}m's method can be even several orders of magnitude  faster than the overlap approach.

Similar to the bosonic analysis in \cite{AGBKa} we consider the non-equilibrium dynamics in two interesting and
experimentally relevant situations. In the QNC setup the application of the Bragg pulse (we
use the Bragg pulse modeling introduced in \cite{RCB,BWENKC} produces a nonsymmetric  momentum  distribution which
is a distinct feature of anyonic systems \cite{HZC1,SC,delC,Patu1}. In \cite{BWENKC}, which treated the bosonic case,
two different time scales were discovered: one of rapid trap-insensitive relaxation right after the application of
the pulse followed by slow periodic behavior. This separation of time-scales is also present in the anyonic system.  During
the periodic behavior the momentum distribution presents alternatively fermionic and anyonic characteristics. This
phenomenon which we call ``periodic dynamical fermionization" is even more pronunced in the dynamics of the anyonic
gas after a quantum quench of the trapping frequency which produces  breathing-mode oscillations. In this
case we also observe another many-body collective effect similar to the one discovered in  \cite{ABGKb} for the bosonic
Tonks-Girardeau gas which is characterized by an additional narrowing of the momentum distribution when the gas is
maximally compressed. In  the anyonic case this narrowing decreases as the statistics parameter increases ($\kappa=0$
for bosons and $\kappa=1$  for fermions) and eventually disappears for the fermionic system.

The plan of the paper is as follows. In Secs. \ref{s1} and \ref{s2} we introduce the anyonic Tonks-Girardeau model,
its eigenfunctions and the one-particle RDM. The derivation of the anyonic generalization of Lenard's formula
is presented in Sec. \ref{s3} and the details of the two numerical methods can be found in Sec. \ref{s4}. The
non-equilibrium dynamics in the QNC setup and after a quantum quench of the trap frequency
is studied in Secs. \ref{s5} and \ref{s6}. We conclude in Sec. \ref{s7}. Some information about the Fredholm minors,
the RDMs of free fermions and two useful theorems are presented in Appendices \ref{a1}, \ref{a2} and \ref{a3}.
A detailed comparison of the two numerical methods used in this paper can be found in Appendix \ref{a4}.

\section{The anyonic Tonks-Girardeau gas}\label{s1}

We consider a one dimensional system of $N$ anyons interacting via a repulsive $\delta$-function potential
in the presence of a confining time-dependent external potential. The second quantized Hamiltonian is
\be\label{ham2}
\mathcal{H}=\int dx\,   \frac{\hbar^2}{2 m}(\6_x\Psi^\dagger)(\6_x\Psi)
+g \Psi^\dagger\Psi^\dagger\Psi\Psi+ V(x,t)\Psi^\dagger\Psi\, ,
\ee
with the anyonic fields $\Psi^\dagger(x), \Psi(x)$ satisfying the following commutation relations
\begin{subequations}\label{comm}
\begin{align}
\Psi(x)\Psi^\dagger(y)&=e^{- i \pi \kappa \varepsilon(x-y)}\Psi^\dagger(y)\Psi(x)+\delta(x-y)\, ,\\
\Psi(x)\Psi(y)&=e^{i \pi \kappa \varepsilon(x-y)}\Psi(y)\Psi(x)\, ,
\end{align}
\end{subequations}
with $\kappa\in [0,1]$ the statistics parameter and $\varepsilon(x)=|x|/x\, ,\varepsilon(0)=0$. Varying the
statistics parameter $\kappa$  in the $[0,1]$ interval  and for $x\ne y$  the anyonic commutation  relations (\ref{comm})
interpolate continuously  between the canonical commutation relations  for bosons ($\kappa=0$) and the
canonical anticommutation  relations for fermions ($\kappa=1$). At coinciding points, $x=y$,
the commutation relations are bosonic in nature. We should point out that an equally valid
choice of interval for the variation of the statistics parameter is given by $\kappa\in[-1,0]$ with the
fermionic system described by $\kappa=-1$. In (\ref{ham2}) $\hbar$ is the reduced Planck constant, $m$ is
the mass of the particles, $g$  quantifies the strength of the repulsive interaction and $V(x,t)$ is the
time-dependent external potential. We are going to mainly consider the case of the parabolic confining
potential with time-dependent frequency $V(x,t)=m \omega^2(t) x^2/2$ but the considerations of this section are also valid
in the case of more complicated time-dependent external potentials.

When $V(x,t)=0$ the Hamiltonian (\ref{ham2}) describes the integrable anyonic Lieb-Liniger model \cite{Kundu,
BGO,PKA} which is the natural generalization to arbitrary statistics of the bosonic Lieb-Liniger model \cite{LL63}.
Introducing the Fock vacuum $|0\rangle$ defined by $\Psi(x)|0\rangle=\langle 0|\Psi^\dagger(x)=0$ for all $x$ and
$\langle 0|0\rangle=1$ the eigenstates of the Hamiltonian (\ref{ham2}) are
\begin{align}
|\boldsymbol{\psi}_N(t)\rangle=&\frac{1}{\sqrt{N!}}\int dz_1 \cdots dz_N\,  \psi_{N,A}(z_1,\cdots,z_N|\,t)\nonumber\\
&\ \times  \Psi^\dagger(z_N,t)\cdots\Psi^\dagger(z_1,t)|0\rangle\, ,
\end{align}
with $\Psi^\dagger(z,t)=e^{i \mathcal{H} t}\Psi^\dagger(z) e^{-i \mathcal{H} t}$. The $N$-body anyonic
wave-function $\psi_{N,A}$  satisfies
\begin{align}\label{asymm}
\psi_{N,A}(\cdots,z_i,z_{i+1},\cdots|\,t)=&e^{i\pi\kappa\varepsilon(z_i-z_{i+1})} \nonumber\\
&\ \times \psi_{N,A}(\cdots, z_{i+1},z_i,\cdots|\,t)\, .
\end{align}
This shows that the wave-functions is symmetric under the  permutation of two particles when the
system is bosonic $(\kappa=0)$ and anti-symmetric when the system is fermionic $(\kappa=1)$. For
an anyonic system, $\kappa\in(0,1)$, the previous relation reveals the broken
space-reversal symmetry characteristic of 1D anyons which results in a non-symmetric momentum distribution.

The first quantized version of (\ref{ham2}) is
\begin{align}\label{ham1}
H=\sum_{i=1}^N\left[-\frac{\hbar^2}{2m}\frac{\6^2}{\6 {z_i}^2}+V(z_i,t)\right]+2 g \sum_{1\le i<j\le N}\delta(z_i-z_j)\, ,
\end{align}
and in this paper we are going to consider the Tonks-Girardeau limit ($g\rightarrow\infty$) of the Hamiltonian
(\ref{ham1}) which imposes and additional hard-core constraint on the wave-function of the anyonic system
$\psi_{N,A}(\cdots, z,\cdots,z,\cdots)|\, t)=0$. In this limit  the anyonic system described by (\ref{ham1}) can be investigated by considering a dual system
of $N$ free fermions described by the Hamiltonian
\begin{align}\label{hamf}
H_F=\sum_{i=1}^N\left[-\frac{\hbar^2}{2m}\frac{\6^2}{\6 {z_i}^2}+V(z_i,t)\right]\, ,
\end{align}
and the wave-functions  can be determined employing the Anyon-Fermi mapping \cite{Girar,PKA1} [$\boldsymbol{z}=(z_1,\cdots,z_N)$]:
\be\label{AFmap}
\psi_{N,A,\boldsymbol{\nu}}(\boldsymbol{z}|\, t)=A(\boldsymbol{z})B(\boldsymbol{z}) \psi_{N,F,\boldsymbol{\nu}}(\boldsymbol{z}|\, t)\, ,
\ee
where
\be
A(\boldsymbol{z})=\prod_{j<k}e^{i \frac{\pi \kappa}{2}\varepsilon(z_j-z_k)}\, ,\
B(\boldsymbol{z})=\prod_{j>k} \varepsilon(z_j-z_k)\, ,
\ee
and $\psi_{N,F,\boldsymbol{\nu}}(\boldsymbol{z}|\, t)$ are the wave-functions of the dual fermionic system. In (\ref{AFmap})
the wave-functions also depend on $\boldsymbol{\nu}=(\nu_1,\cdots,\nu_N)$ which identify the  single-particle energy levels.
The fermionic wave-functions are Slater determinants (which are eigenstates of (\ref{hamf}) and constitute a basis of the Fock space) of the single-particle (SP) wavefunctions
\be\label{Slater}
\psi_{N,F,\boldsymbol{\nu}}(\boldsymbol{z}|\, t)=\frac{1}{\sqrt{N!}}\mbox{det}_{i,j=1}^N \phi_{\nu_i}(z_j,t)\, ,
\ee
where $H_{\mbox{\tiny{SP}}}(z,0)\phi_{\nu_i}(z)=E_{\nu_i}\phi_{\nu_i}(z)$ and $i\hbar \6 \phi_{\nu_i}(z,t)/\6 t=H_{\mbox{\tiny{SP}}}(z,t) \phi_{\nu_i}(z,t)$
with $H_{\mbox{\tiny{SP}}}(z,t)=-(\hbar^2/2m)(\6^2/\6 z^2)+V(z,t)$. At $t=0$ the energy of the $N$-body state (\ref{Slater}) is
$E_{N,\boldsymbol{\nu}}=\sum_{i=1}^N E_{\nu_i}.$

We should point out that the Anyon-Fermi mapping remains valid even in the case of a general potential energy
as long as it includes a hard-core of radius $a\ge 0$ \cite{Len66,PKA1}. In the case of hard-wall boundary conditions
or when the systems are subjected to an confining external potential the same boundary conditions hold for
the anyonic and fermionic systems and the energy eigenvalues are equal. In the case of periodic or twisted boundary
conditions the situation is more complicated (see \cite{Len66,PKA1}).

\section{One-body reduced density matrix}\label{s2}

In this paper we are interested in investigating the dynamics of the anyonic one-body reduced density
matrix  at finite temperature defined by [$\boldsymbol{\bar{z}}=(z_1, \cdots,z_{N-1})$]
\begin{align}\label{reduced}
\rho^{(1)}(x,y|\, t)=&\sum_{N=1}^\infty\sum_{\boldsymbol{\nu}}p(N,\boldsymbol{\nu}) N\int dz_1\cdots dz_{N-1}\nonumber\\
&\ \times\psi_{N,A,\boldsymbol{\nu}}(\boldsymbol{\bar{z}},x|\,t) \psi_{N,A,\boldsymbol{\nu}}^*(\boldsymbol{\bar{z}},y|\,t)\, ,
\end{align}
with $p(N, \boldsymbol{\nu})$ the probabilities of an arbitrary statistical ensemble. Following \cite{AGBKa} (see also
\cite{XR17} for the lattice case) we consider the system initially in thermal equilibrium
described by  the grand-canonical ensemble with $p(N,\boldsymbol{\nu})=e^{-(E_{N,\boldsymbol{\nu}}
-\mu N)/k_B T_0}/\mathcal{Z}$, $\mu$ the chemical potential and $T_0$ the equilibrium temperature at
$t=0$. $\mathcal{Z}=\sum_{N,\boldsymbol{\nu}} e^{-(E_{N,\boldsymbol{\nu}}-\mu N)/k_B T_0}$ is the grand-canonical
partition function and $\psi_{N,A,\boldsymbol{\nu}} (\boldsymbol{\bar{z}},x|\,t)$ are the evolved wave-functions
obtained from $\psi_{N,A,\boldsymbol{\nu}} (\boldsymbol{\bar{z}},x|\,0)$ with the Hamiltonian (\ref{ham1}).
From the one-body density matrix we can obtain two very important and experimentally accessible quantities: the
real space density $\rho(x,t)=\rho^{(1)}(x,x|\,t)$ and the momentum distribution
\be
n(k,t)=\int\int e^{-i k(x-y)} \rho^{(1)}(x,y|\,t)\, dxdy\, .
\ee
We should point out that while the real-space density is independent of the statistics (this can be seen easily from the
Anyon-Fermi mapping (\ref{AFmap}) and (\ref{reduced})) the momentum distribution is highly dependent on $\kappa$.

\section{Anyonic generalization of Lenard's formula}\label{s3}

In the 60's Lenard used the Bose-Fermi mapping to derive  an expansion of the bosonic reduced density
matrices in terms of the fermionic reduced matrices \cite{Len66}. Lenard's result is valid not only for
the one-body RDM  but also for the more general case of the $n$-body density matrices defined by
\begin{align}\label{reducedn}
\rho^{(n)}(\boldsymbol{x},\boldsymbol{y}|\, t)=&\sum_{N=n}^\infty\sum_{\boldsymbol{\nu}}p(N,\boldsymbol{\nu}) \frac{N!}{(N-n)!}\int dz_1\cdots dz_{N-n}\nonumber\\
&\ \times\psi_{N,A,\boldsymbol{\nu}}(\boldsymbol{\tilde{z}},\boldsymbol{x}|\,t) \psi_{N,A,\boldsymbol{\nu}}^*(\boldsymbol{\tilde{z}},\boldsymbol{y}|\,t)\, ,
\end{align}
with $\boldsymbol{x}=(x_1,\cdots,x_n)\, ,\boldsymbol{y}=(y_1,\cdots,y_n)$ and $\boldsymbol{\tilde{z}}=(z_{1},\cdots,z_{N-n})$.
The anyonic generalization of Lenard's formula was derived in \cite{PKA1}. In order to make the paper self-contained here
we rederive it in the particular case of the one-body density matrix. We will need a simple result \cite{Len66,PKA1}:
\begin{widetext}
\begin{lem}\label{lemma}
For any symmetric function $f(z_1,\cdots,z_n)$, a constant $\xi$  and $I$ an interval in the domain $\Omega$ we have
\be
\int_\Omega dz_1\cdots\int_{\Omega} dz_n\, \xi^{\sigma(I)} f(z_1,\cdots,z_n)=\sum_{j=0}^n C_j^n (-1+\xi)^j
\int_I dz_1\cdots\int_I dz_j\int_\Omega dz_{j+1}\cdots\int_{\Omega} d z_n\, f(z_1,\cdots,z_n)\, ,
\ee
where $\sigma(I)$ counts the number of variables $z_1,\cdots,z_n$ contained in $I$.
\end{lem}
We start from the definition of the reduced density matrix (\ref{reduced}). Using the Anyon-Fermi mapping (\ref{AFmap}) for the wave-functions we have
\begin{align}
\psi_{N,A,\boldsymbol{\nu}}(\boldsymbol{\bar{z}},x|\,t) \psi_{N,A,\boldsymbol{\nu}}^*(\boldsymbol{\bar{z}},y|\,t)
=&\prod_{1\le j\le N-1}\left( e^{i\frac{\pi\kappa}{2}[\varepsilon(z_j-x)-\varepsilon(z_j-y)] }\varepsilon(x-z_j)\varepsilon(y-z_j)\right)
\psi_{N,F,\boldsymbol{\nu}}(\boldsymbol{\bar{z}},x|\,t) \psi_{N,F,\boldsymbol{\nu}}^*(\boldsymbol{\bar{z}},y|\,t)\, ,\nonumber\\
=&\left(-e^{i\pi\kappa\varepsilon(y-x)}\right)^{\sigma(I)} \psi_{N,F,\boldsymbol{\nu}}(\boldsymbol{\bar{z}},x|\,t) \psi_{N,F,\boldsymbol{\nu}}^*(\boldsymbol{\bar{z}},y|\,t)\, ,
\end{align}
with $\boldsymbol{\bar{z}}=(z_1, \cdots,z_{N-1})$ and $I=[x,y]$ when $y>x$ and $I=[y,x]$ when $x>y$. Introducing $\xi=-e^{i\pi\kappa\varepsilon(y-x)}$
and plugging the previous result in (\ref{reduced}) followed by the application of the Lemma \ref{lemma} we find
\begin{align}
\rho^{(1)}(x,y|\, t)=&\sum_{N=1}^\infty\sum_{\boldsymbol{\nu}}p(N,\boldsymbol{\nu}) N\int dz_1\cdots dz_{N-1}\,
\xi^{\sigma(I)}\psi_{N,F,\boldsymbol{\nu}}(\boldsymbol{\bar{z}},x|\,t) \psi_{N,F,\boldsymbol{\nu}}^*(\boldsymbol{\bar{z}},y|\,t)\, ,\nonumber\\
=& \sum_{N=1}^\infty\sum_{\boldsymbol{\nu}}p(N,\boldsymbol{\nu}) N
\left(
 \sum_{j=0}^{N-1} C_j^{N-1} (-1+\xi)^j \int_I dz_1\cdots\int_I dz_j\int dz_{j+1}\cdots\int dz_{N-1}
 \psi_{N,F,\boldsymbol{\nu}}(\boldsymbol{\bar{z}},x|\,t) \psi_{N,F,\boldsymbol{\nu}}^*(\boldsymbol{\bar{z}},y|\,t)
 \right)\, .\nonumber
\end{align}
In the last relation changing the order of the summation we can identify in the right hand side the reduced density
matrices of free fermions, denoted by $\rho_F^{(n)}$, (for an alternative definition equivalent with (\ref{reducedn}) in the case of fermions see (\ref{reducednf}))
obtaining
\begin{align}\label{i1}
\rho^{(1)}(x,y|\, t) =& \sum_{j=0}^\infty \frac{(-1+\xi)^j}{j!}\int_I dz_1\cdots\int_I dz_j\, \rho_F^{(j+1)}(z_1,\cdots,z_j,x;z_1,\cdots,z_j,y|\, t)\, ,\nonumber\\
=& \sum_{j=0}^\infty \frac{(-1+\xi)^j}{j!}[\varepsilon(y-x)]^j \int_x^y dz_1\cdots\int_x^y dz_j\, \rho_F^{(j+1)}(z_1,\cdots,z_j,x;z_1,\cdots,z_j,y|\, t)\, .
\end{align}
The reduced density matrices of free fermions are computed in Appendix~\ref{a2}. Using  formula (\ref{redff}) in (\ref{i1}) we obtain the main result of this section:
the anyonic generalization of Lenard's formula for the one-body RDM is
\be\label{lenard}
\rho^{(1)}(x,y|\, t) =\sum_{j=0}^\infty\frac{(-\gamma)^j}{j!}\int_x^y dz_1\cdots\int_x^y
\rho_F^{(1)}\left(\begin{array} {cccc}
                            x &z_1&\cdots& z_j\\
                            y &z_1&\cdots& z_j
                   \end{array}; t\right)\, ,   \ \
\gamma=\varepsilon(y-x)\left (1+e^{i\pi\kappa\varepsilon(y-x)}\right)\, ,
\ee
\end{widetext}
where we have used the notation introduced in (\ref{ai2}) and $\rho_F^{(1)}(x,y|\, t)=\sum_{i=0}^\infty
f_i \phi_i(x,t)\phi_i^*(y,t)$  with $f_i=[e^{(E_i-\mu)/k_B T_0}+1]^{-1}$   the Fermi-Dirac occupation factor.
The $j=0$ term  of (\ref{lenard}) is given by $\rho_F^{(1)}(x,y|\, t)$ which is the one-body reduced
density matrix of the dual fermionic system.

\section{Fredholm minor representation for the reduced density matrix}\label{s4}

While it is useful to derive the short distance expansion of the one-body RDM \cite{FFGW03a,FFGW03b,VM}
the numerical implementation of Lenard's formula (\ref{lenard}) is computationally involved even if we truncate the
series after the first few terms. The efficient calculation of  physical relevant quantities, such as the momentum
distribution, and not only for large values of $k$, requires a more computationally friendly representation of (\ref{lenard}).
A more fruitful approach is to realize that (\ref{lenard}) is the first Fredholm minor (see formula (\ref{ai4})) of
an integral operator $1-\gamma \hat\rho_F^{(1)}$ with kernel $\rho_F^{(1)}(\lambda,\mu|\,t)$ which acts on arbitrary
function like $\int_x^y \rho_F^{(1)}(\lambda,\mu|\,t)f(\mu)\, d\mu$ and  $\gamma$ is defined in
(\ref{lenard}) \cite{Len66,FFGW03a,FFGW03b}. Introducing the resolvent kernel of this integral operator which satisfies
\be\label{resolventi}
\textsf{R}_F(\lambda,\mu|\, t)=\rho_F^{(1)}(\lambda,\mu|\, t)+
\gamma\int_{x}^y\rho_F^{(1)}(\lambda,\nu|\, t)\textsf{R}_F(\nu,\mu|\, t)\, d\nu\, ,
\ee
and using the identity (\ref{ai6}) we find
\be\label{fred}
\rho^{(1)}(x,y|\, t)=\textsf{R}_F(x,y|\, t)\, \mbox{\textsf{det}}(1-\gamma\hat{\rho}_F^{(1)})\, .
\ee
The representation (\ref{fred}) for arbitrary statistics was first derived in \cite{PKA1}. An efficient numerical evaluation of (\ref{fred}) (or, equivalently of (\ref{lenard})) in the case of impenetrable bosons at
zero temperature was  proposed by Pezer and Buljan in  \cite{PB07} and in the case of arbitrary statistics by del Campo in \cite{delC}.
The finite temperature generalization for impenetrable bosons was derived by Atas \textit{et al.} in  \cite{AGBKa}.
In this paper we are going to use another approach based on the numerical evaluation of the resolvent and Fredholm determinant
using Nystr\"{o}m's method \cite{PFTV,Born}. In addition, we will also derive the anyonic generalization of the
result obtained in \cite{AGBKa}. In Appendix~\ref{a4} we show that when there is no analytical formula for the overlaps of
the single-particle wave-functions (which is the case in many experimentally relevant cases such as the QNC
setup) this approach significantly outperforms the methods  of \cite{PB07} and \cite{AGBKa} for moderate and large number of
particles at zero and finite temperatures.

The starting point of both methods is represented by the truncation of the sum which describes the free fermionic one-body density
matrix after the first $M$ terms
\be\label{truncation}
\rho_F^{(1)}(x,y|\, t)\simeq\sum_{i=0}^{M-1}f_i \phi_i(x,t)\phi_i^*(y,t)\, .
\ee
The truncation
parameter $M$ has to be chosen such that the discarded terms in the infinite sum are negligible. At zero temperature we have
$M=N$ with $N$ the number of particles in the system and $\rho_F^{(1)}(x,y|\, t)=\sum_{i=0}^{N-1} \phi_i(x,t)\phi_i^*(y,t)$.
It is also assumed that the SP wave-functions $\{\phi_i(x,t)\}_{i=0}^{M-1}$ are known either analytically or by numerically
solving the time-dependent  Schr\"odinger equation.

\subsection{Numerical evaluation of the RDM using Nystr\"{o}m's method}\label{s42}

Assuming that we know the one-body RDM of free fermions $\rho_F^{(1)}(\lambda,\mu|\, t)$ probably the simplest method of evaluating (\ref{fred})
is  by solving the integral equation satisfied by the resolvent (\ref{resolventi}) followed by the evaluation
of the Fredholm determinant. Eq.~(\ref{resolventi}) is a Fredholm integral equation of the second kind with a smooth kernel
which can be very easily and accurately solved using Nystr\"om's method (Chap. 18 of \cite{PFTV}) and, as it was shown in
\cite{Born}, the same method can  be used to evaluate the Fredholm determinant. Briefly, the necessary steps are the following.
Consider a quadrature (for a pragmatic introduction in numerical integration with quadratures see Chap. 4 of \cite{PFTV}) with
positive weights which approximates the integral over $[x,y]$ of an arbitrary function $f(\lambda)$ (reasonably well-behaved) as
\be\label{i20}
\int_x^y f(\lambda) d\lambda \simeq \sum_{j=1}^m w_j f(\lambda_j)\, .
\ee
In (\ref{i20})  $\{w_j\}_{j=1}^m$ are called the quadrature weights and  $\{\lambda_j\}_{j=1}^m$ the quadrature points or abscissas.
Using this quadrature the integral equation (\ref{resolventi}) for $\mu=y$ can be written as
\begin{align}\label{resolventinterp}
\textsf{R}_F(\lambda,y|\, t)&=\rho_F^{(1)}(\lambda,y|\, t)\nonumber\\
&\ \ \ \  +\gamma\sum_{j=1}^m w_j \rho_F^{(1)}(\lambda,\lambda_j|\, t)\textsf{R}_F(\lambda_j,y|\, t)\, .
\end{align}
Considering (\ref{resolventinterp}) at all points of the quadrature $\{\lambda_j\}_{j=1}^m$ we obtain the
following system of $m$ linear equations
\be\label{i21}
\left(1-\gamma{\bar{\rho}}_F^{(1)}\right)R=\rho\, ,
\ee
with ${\bar{\rho}}_F^{(1)}$ a square matrix of dimension $m$ and elements $w_j \rho_F^{(1)}(\lambda_i,\lambda_j|\, t)$  and
\begin{align}
R=&\left[\textsf{R}_F(\lambda_1,y|\, t),\cdots,\textsf{R}_F(\lambda_m,y|\, t)\right]^T\, ,\nonumber\\
\rho=&\left[\rho_F^{(1)}(\lambda_1,y|\, t),\cdots,\rho_F^{(1)}(\lambda_m,y|\, t)\right]^T\, .
\end{align}
Then $R=\left(1-\gamma{\bar{\rho}}_F^{(1)}\right)^{-1}\rho$  and $\textsf{R}_F(x,y|\, t)$ can be computed
using the solution for $R$ and (\ref{resolventinterp}) for $\lambda= x$. Therefore, for the  computation of $\textsf{R}_F(x,y|\, t)$
we need only to solve the system of linear equations (\ref{i21}) followed by the use of the interpolation formula (\ref{resolventinterp}).
The evaluation of the Fredholm determinant is even simpler \cite{Born} and is given by
\be\label{i22}
\mbox{\textsf{det}}(1-\gamma\hat{\rho}_F^{(1)})=
\mbox{det}\left(\delta_{ij}-\gamma w_i^{1/2}\rho_F^{(1)}(\lambda_i,\lambda_j|\, t)w_j^{1/2}\right)_{i,j=1}^m\, .
\ee
We make two observations. First, the outlined derivation does not depend explicitly on the truncation level $M$ so in
principle it is also valid if we would know the free fermionic RDM through other methods and not from (\ref{truncation}).
Second, it is also valid at zero temperature.  In this case Eq.~(\ref{truncation}) reduces to its zero-temperature expression.

In summary, the numerical evaluation of the reduced density matrix (\ref{fred}) using Nystr\"{o}m's method consists
of the following steps: a) determination (analytically or numerically) of the SP wave-functions $\{\phi_i(x,t)\}_{i=0}^{M-1}$;
b) solving the linear system (\ref{i21}) with the free fermionic RDM given by (\ref{truncation}); c) determination of
$\textsf{R}_F(x,y|\, t)$ using the solution of the linear system and interpolation formula (\ref{resolventinterp});
d) computation of the Fredholm determinant using (\ref{i22}).

\subsection{Numerical evaluation of the RDM using the truncated basis}\label{s43}

In this section we are going to obtain the generalization to arbitrary statistics of the numerical
method  derived  in  \cite{AGBKa} for impenetrable bosons.
It will be useful to introduce the notation
\be
\tilde{\phi}_i(\lambda,t)=\sqrt{f_{i-1}}\phi_{i-1}(\lambda,t)\, ,\ \ i=1,\cdots,M\, ,
\ee
in terms of which (\ref{truncation}) takes the form  $\rho_F^{(1)}(\lambda,\mu|\, t)=\sum_{i=1}^{M}\tilde{\phi}_i(\lambda,t)\tilde{\phi}_i^*(\mu,t)$.
We want to obtain numerically manageable expressions for the resolvent and Fredholm determinant in the truncated
basis  $\{\tilde{\phi}_i(\lambda,t)\}_{i=1}^{M}$.

\textit{Resolvent kernel in the truncated basis.} Plugging the truncated expression for $\rho_F^{(1)}(\lambda,\mu|\, t)$
in the integral equation satisfied by the resolvent kernel Eq.~(\ref{resolventi}) we obtain
\be\label{i2}
\textsf{R}_F(\lambda,\mu|\, t)=\sum_{j=1}^M\tilde{\phi}_j(\lambda,t)\tilde{\phi}_j^*(\mu,t)+
\gamma\sum_{j=1}^M \tilde{\phi}_j(\lambda,t)A_j(\mu,t)\, ,
\ee
where we have introduced
\[
A_j(\mu,t)=\int_x^y \tilde{\phi}^*_j(\nu,t)\textsf{R}_F(\nu,\mu|\,t)\, d\nu\, .
\]
The $A_j$ coefficients can be determined by multiplying Eq.~(\ref{i2}) with $\tilde{\phi}_j^*(\lambda,t)$
and integrating over the $[x,y]$ interval. We find
\be\label{i3}
A_j(\mu,t)=\sum_{i=1}^M S_{ij}(t)\left[\tilde{\phi}_i^*(\mu,t)+\gamma A_i(\mu,t) \right]\, ,
\ee
with $\boldsymbol{S}(t)$ a square matrix of dimension $M$ with elements
\be
S_{ij}(t)=\int_x^y\tilde{\phi}_i(\lambda,t)\tilde{\phi}_j^*(\lambda,t)\, d\lambda\, .
\ee
We want to rewrite the previous results in a matrix form transparent notation. We introduce two
column vectors defined by $A=(A_1,\cdots,A_M)^T$ and $\tilde{\phi}^*=(\tilde{\phi}_1^*,\cdots,\tilde{\phi}_M^*)^T$.
Then, (\ref{i3}) can be written as $(1-\gamma \boldsymbol{S}^T)\, A=\boldsymbol{S}^T\,\phi^*$ or, equivalently, as
$ A=(1-\gamma \boldsymbol{S}^T)^{-1}\boldsymbol{S}^T\,\phi^*$. From this last relation and Eq.~(\ref{i2}) we see that
\begin{align}\label{resolventt}
\textsf{R}_F(\lambda,\mu|\, t)=&\sum_{i,j=1}^M \tilde{\phi}_i(\lambda,t)\nonumber\\
&\times \left(\delta_{ij}+\gamma\left[(1-\gamma \boldsymbol{S}^T)^{-1}\boldsymbol{S}^T\right]_{ij}\right) \tilde{\phi}_j^*(\mu,t)\, ,\nonumber\\
&=\sum_{i,j=1}^M \tilde{\phi}_i(\lambda,t) (1-\gamma \boldsymbol{S}^T)^{-1}_{ij} \tilde{\phi}_j^*(\mu,t)\, .
\end{align}
which represents the resolvent kernel in the truncated basis.

\textit{Fredholm determinant in the truncated basis.}  Now we are going to show that
in the truncated basis we have
\be\label{tdet}
\mbox{\textsf{det}}(1-\gamma\rho_F^{(1)})=\mbox{det}(1-\gamma \boldsymbol{S})\, .
\ee
We point out that in the left hand side of the previous relation we have a Fredholm determinant (see \ref{ai3})
while on the right hand side we have an usual determinant of a square matrix of dimension $M$.  From the definition of the
Fredholm determinant we have (see (\ref{ai2}) and (\ref{ai3}))
\begin{align}\label{i4}
\mbox{\textsf{det}}(1-\gamma\rho_F^{(1)})=&1+\sum_{n=1}^\infty\frac{(-\gamma)^n}{n!}\int_x^yd\xi_1\cdots\int_x^yd\xi_n\nonumber\\
&\ \ \ \ \times \rho_F^{(1)}\left(\begin{array}{cccc}
              \xi_1  &\cdots & \xi_n\\
              \xi_1 &\cdots  & \xi_n
          \end{array}; t
\right)\, .
\end{align}
On the other hand using the von Koch formula for the determinant we have
\begin{align}\label{i5}
\mbox{det}(1-\gamma \boldsymbol{S})=&1+\sum_{n=1}\frac{(-\gamma^n)}{n!} \sum_{k_1,\cdots,k_n=1}^M
\mbox{det} \left(S_{k_p,k_q}\right)_{p,q=1}^n\, .
\end{align}
We will show that: a) all the terms with $n>M$ in the right hand side of (\ref{i4}) vanish and
b) for all $n\le M$ the terms in the expansions (\ref{i4}) and (\ref{i5}) are equal proving (\ref{tdet}).
Let $n\le M$. Then
\begin{align}
 \rho_F^{(1)}\left(\begin{array}{cccc}
              \xi_1  &\cdots & \xi_n\\
              \xi_1 &\cdots  & \xi_n
          \end{array}; t
\right)=
\left|
\begin{array}{cccc}
        \boldsymbol{a}_1^2    &      \boldsymbol{a}_1\cdot\boldsymbol{a}_2  &   \cdots     &\boldsymbol{a}_1\cdot\boldsymbol{a}_n\\
        \boldsymbol{a}_2\cdot\boldsymbol{a}_1   &   \boldsymbol{a}_2^2      &      \cdots     &    \boldsymbol{a}_2\cdot\boldsymbol{a}_n\\
        \vdots & \vdots & \ddots  & \vdots\\
        \boldsymbol{a}_n\cdot \boldsymbol{a}_1  & \boldsymbol{a}_n\cdot \boldsymbol{a}_2   &\cdots    & \boldsymbol{a}_n^2
        \end{array}
\right|\nonumber
\end{align}
with $\boldsymbol{a}_i=(\tilde{\phi}_1(\xi_i,t),\cdots,\tilde{\phi}_M(\xi_i,t))$  and scalar product
$\boldsymbol{a}_i\cdot \boldsymbol{a}_j=\sum_{k=1}^M a_{ik} a_{jk}^*$. From this representation we see that
the only nonvanishing terms in the expansion are the ones with $n\le M$ (if $n>M$ we would have on the
right hand side the Gram determinant of a number of vectors which is larger than the
dimension of the linear space and therefore is equal to zero).  Using  Thm.~(\ref{thm1}) of  Appendix \ref{a3} we obtain
\begin{align}\label{i6}
 \rho_F^{(1)}&\left(\begin{array}{cccc}
              \xi_1  &\cdots & \xi_n\\
              \xi_1 &\cdots  & \xi_n
          \end{array}; t
\right)=\nonumber\\
&\frac{1}{n!}\sum_{k_1,\cdots,k_n=1}^M
\left|
\begin{array}{cccc}
        a_{1k_1} & a_{1k_2} &\cdots &a_{1k_n}\\
        a_{2k_1} & a_{2k_2} &\cdots &a_{2k_n}\\
        \vdots & \vdots & \ddots  & \vdots\\
         a_{nk_1} & a_{nk_2} &\cdots &a_{nk_n}\\
        \end{array}
\right|^2\, ,
\end{align}
with $a_{ik}=\tilde{\phi}_k(\xi_i,t)$. In the case of the von Koch expansion (\ref{i5}) employing Thm. \ref{thm2} of Appendix \ref{a3}
we have
\begin{align}\label{i7}
&\sum_{k_1,\cdots,k_n=1}^M \mbox{det} \left(S_{k_p,k_q}\right)_{p,q=1}^n
=\frac{1}{n!}\sum_{k_1,\cdots,k_n=1}^M\nonumber\\
&\ \ \ \ \ \ \  \int_x^y d\xi_1\cdots\int_x^y d\xi_n \left|
\begin{array}{cccc}
a_{1 k_1} &a_{2 k_1} &\cdots& a_{n k_1}\\
a_{1 k_2} &a_{2 k_2} &\cdots& a_{n k_2}\\
\vdots & \vdots & \ddots  & \vdots\\
a_{1 k_n} &a_{2 k_n} &\cdots& a_{n k_n}\\
\end{array}
\right|^2\, .
\end{align}
Then (\ref{tdet}) follows from (\ref{i7}) and (\ref{i6}) and the fact that $\mbox{det} A=\mbox{det} A^T$
for an arbitrary square matrix $A$.  An alternative derivation of this result can be obtained using Plemelj's
formula for the Fredholm determinant as in \cite{CMV11}.

Plugging (\ref{resolventt}) and (\ref{tdet}) in (\ref{fred}) and using $(1-\gamma S^T)^{-1}=[(1-\gamma S)^{-1}]^T$
we obtain
\be\label{rhot}
\rho^{(1)}(x,y|\, t)=\sum_{i,j=0}^{M-1}\sqrt{f_i}\phi_i(x,t)Q_{ij}(x,y|\,t)\sqrt{f_j}\phi_j^*(y,t)\, ,
\ee
with
\be\label{i8}
\boldsymbol{Q}(x,y|\, t)=[\boldsymbol{P}^{-1}(x,y|\, t)]^T\mbox{det}\, \boldsymbol{P}(x,y|\,t)\, ,
\ee
In (\ref{rhot}) and (\ref{i8}) $\boldsymbol{Q}$ and $\boldsymbol{P}$ are square matrices of dimension
$M$ with indices $i,j=0,\cdots,M-1$ and the elements of $\boldsymbol{P}$ are
\be\label{defP}
P_{ij}(x,y|\, t)=\delta_{ij}-\gamma \sqrt{f_i f_j} \int_x^y \phi_i(\lambda,t)\phi_j^*(\lambda,t)\, d\lambda\, ,
\ee
with $\gamma=\varepsilon(y-x)\left (1+e^{i\pi\kappa\varepsilon(y-x)}\right)$.

In the bosonic case ($\kappa=0$)
(\ref{rhot}) reproduces the result from \cite{AGBKa} and for $\kappa=1$ it reduces to
$\rho^{(1)}(x,y|\, t)=\sum_{i=0}^{M-1}f_i\phi_i(x,t)\phi_j^*(y,t)\, $ which is just the free fermionic
result as expected. At zero temperature the occupation factors become zero for $M>N-1$ and we obtain
the results derived in \cite{PB07,delC}. The necessary
steps required to compute the RDM of an anyonic system using the method presented in this section are the
following: a) determination of the SP wave-functions; b) calculation of the wave-functions overlaps and
of the matrix $\boldsymbol{P}$ using (\ref{defP}); c) computation of the matrix $\boldsymbol{Q}$ (\ref{i8});
d) performing the summation in (\ref{rhot}).

\subsection{Comparison of the two methods}

A detailed analysis of the complexity of the two methods outlined above can be found in Appendix~\ref{a4}. We have
two distinct situations. If the overlaps of the wave-functions $\int_x^y \phi_i(\lambda,t)\phi_j^*(\lambda,t)\, d\lambda$
can be analytically computed then the truncated basis method (\ref{rhot}) is the method of choice and will
be used in Section \ref{s6}. If the overlaps are not amenable to analytical calculations and numerical integration is
required then Nystr\"{o}m's method is preferable with the exception of the case of small number of particles (or truncation level $M$)
and large $|x-y|$. In particular, Nystr\"{o}m's method is extremely efficient, especially at finite temperatures
(see Appendix~\ref{a4}), because: a) $M$ increases with temperature and b) due to the exponential decay the RDM is concentrated
in a narrow strip $|x-y|<C $ with the contributions from $|x-y|>C$  being negligible.

\section{Anyonic quantum Newton's cradle}\label{s5}

\begin{figure*}
\includegraphics[width=1\linewidth]{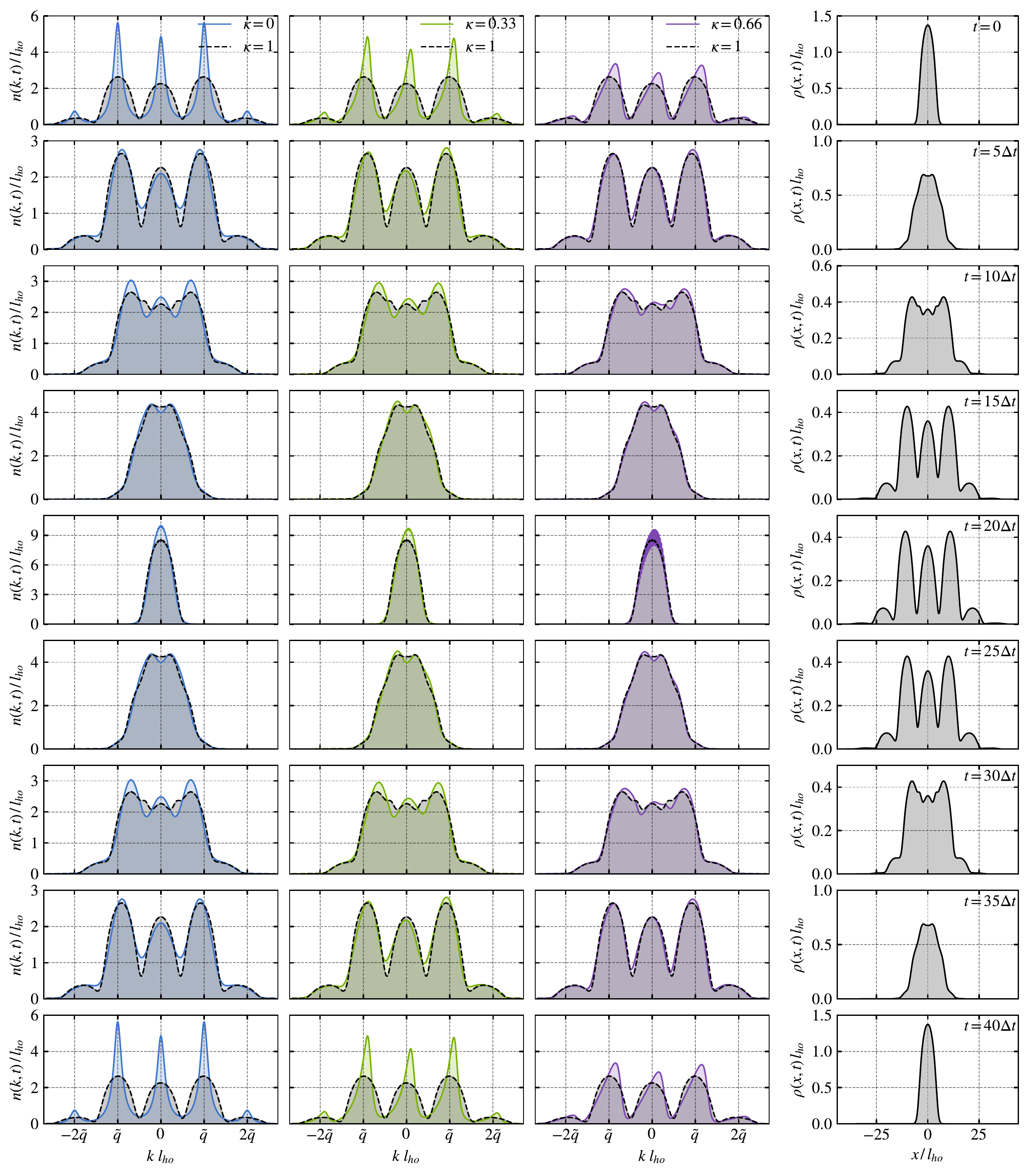}
\caption{Dynamics of the momentum distribution  of impenetrable anyons ($\kappa=0$ first column, $\kappa=0.33$ second column,
and $\kappa=0.66$ third column) and free fermions ($\kappa=1$) in the QNC setup. The last column presents the dynamics of the real-space density
which is independent of statistics. Here $N=10$, $A=1.5$, $q l_{ho}=10.73$ ($q=6\pi$, $l_{ho}=0.57$, $\omega=3$), and the initial dimensionless temperature is $\theta_0=0.083$ ($T_0=0.25$).
The time step is $\Delta t=\pi/(40 \omega)$ and $\tilde {q}=q l_{ho}$.}
\label{NCall}
\end{figure*}

As a first application of the formalism developed in the previous sections we are studying the dynamics of
the anyonic TG gas in the QNC setting at finite temperature. In the original experiment
\cite{KWW07} a quasi-1D gas of $^{87}$Rb atoms in a weakly harmonical potential in the longitudinal direction
was subjected to a sequence of Bragg pulses designed to split the system in two counter propagating halves.
After a short dephasing  period the atoms continued to collide repeatedly for hundreds of times without
thermalizing as in the  case of a 3D system. Such long lived non-thermal states are called pre-thermal and
they are a consequence of the near-integrability of the system  under consideration. Recent realizations of
similar or slightly modified setups involving dipolar dysprosium atoms were reported  in \cite{LZMSZ,TKLSM}.
Theoretical and numerical  investigations of impenetrable bosons in the QNC setup using the Quench Action
\cite{CE,Caux1}  and the numerical method presented in Sec. \ref{s43} were previously performed both at zero
\cite{BWENKC}  and  finite  temperature \cite{AGBKa}. A comprehensive numerical treatment of the finite
coupling case using the generalized hydrodynamics \cite{CDY,BCNF} can be found in \cite{CDDKY} (see also \cite{SBDD19}).

For the modeling of the Bragg pulse which initiates the oscillation we are going to use the results of
\cite{RCB,BWENKC}. In the notation and terminology of the latter reference we are going to consider the case
of the so-called Kapitza-Dirac pulse which can be described by the operator ($\hbar=k_B=1$)
\be\label{bragg}
U(q,A)=\exp\left(- i A\int \cos(q z)\, \Psi^\dagger(z)\Psi(z) \,  dz\right)\, .
\ee
For any state of the system $\bm{\psi}\rangle$ the effect of this instantaneous pulse is to produce a new state
$|\bm{\psi}_{q,A}\rangle=U(q,A)|\bm{\psi}\rangle$. In typical experiments $A\sim 1$ and $q\sim 2\pi n$ with $n$
the density.  The long-pulse Bragg regime can be modeled as in \cite{AGBKa} by adding to the harmonic potential
a periodic lattice potential $V_B(z,t)=\Omega(t)\cos(2 q z)$ with  $\Omega(t)$ a sequence of two square pulses.

The evolution of the system after the Bragg pulse is driven by the Hamiltonian (\ref{ham1}) with $g=\infty$.
The dual fermionic system describes free fermions in a harmonic potential  $V(z,t)=m \omega^2 z^2/2$ for which the
single-particle   wave-functions are the well known Hermite functions
\be\label{hermite}
\phi_j(z)=e^{- m\omega^2 z^2/2}\frac{1}{\sqrt{2^j j!}} \left(\frac{m \omega}{\pi}\right)^{1/4}H_j(\sqrt{mw}\, z)\, ,
\ee
with $H_j(z)$ denoting the Hermite polynomials and $E_j= \omega(j+1/2).$  The harmonic oscillator length is $l_{ho}=\sqrt{1/m \omega}$.
The action of the Bragg pulse (\ref{bragg})
on the SP wave-functions is given by $U(q,A)\phi_j(z)=e^{-i A \cos(q z)}\phi_j(z)$. The time evolved wave-functions
can be determined using the propagator of the quantum harmonic oscillator \cite{BWENKC}
\be
\phi_j(z,t)=\inti K(z,u|\, t)e^{-i A \cos (q u)}\phi_j(u)\, du\, ,
\ee
with (see 2.5.18 of \cite{Sakurai})
\begin{align}
K(z,u|\, t)&=\left(\frac{m \omega}{2\pi i \sin(\omega t)}\right)^{1/2}\nonumber\\
&\times\exp\left(\frac{-m\omega(z^2+u^2)\cos(\omega t)+2 m\omega z u}{2 i\sin(\omega t)}\right)\, .
\end{align}

\begin{figure}
\includegraphics[width=1\linewidth]{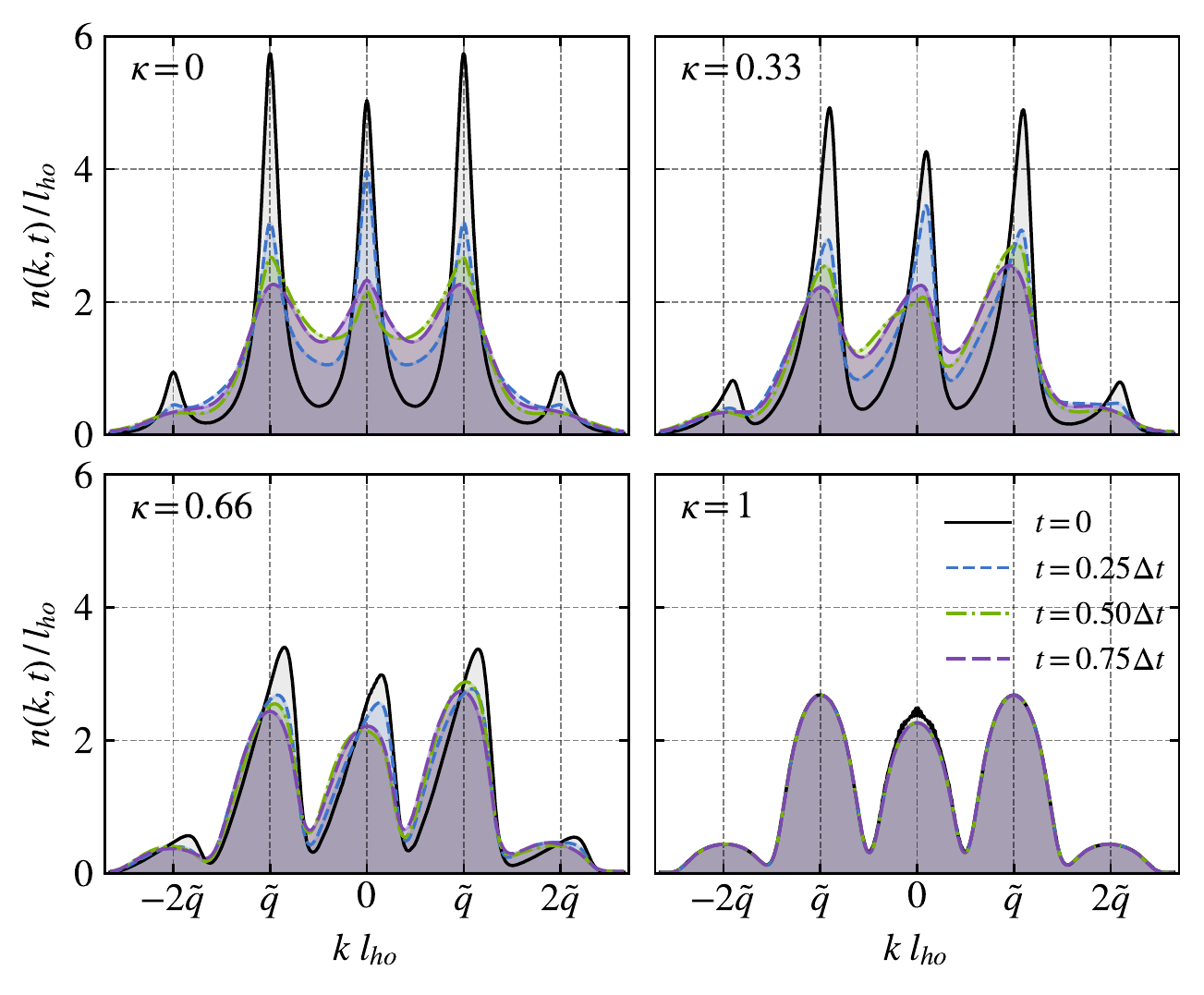}
\caption{Evolution of the momentum distribution immediately after the pulse in the QNC setup. The  parameters of the system are the
same as in Fig.~\ref{NCall}.
 }
\label{First}
\end{figure}
\begin{figure}
\includegraphics[width=1\linewidth]{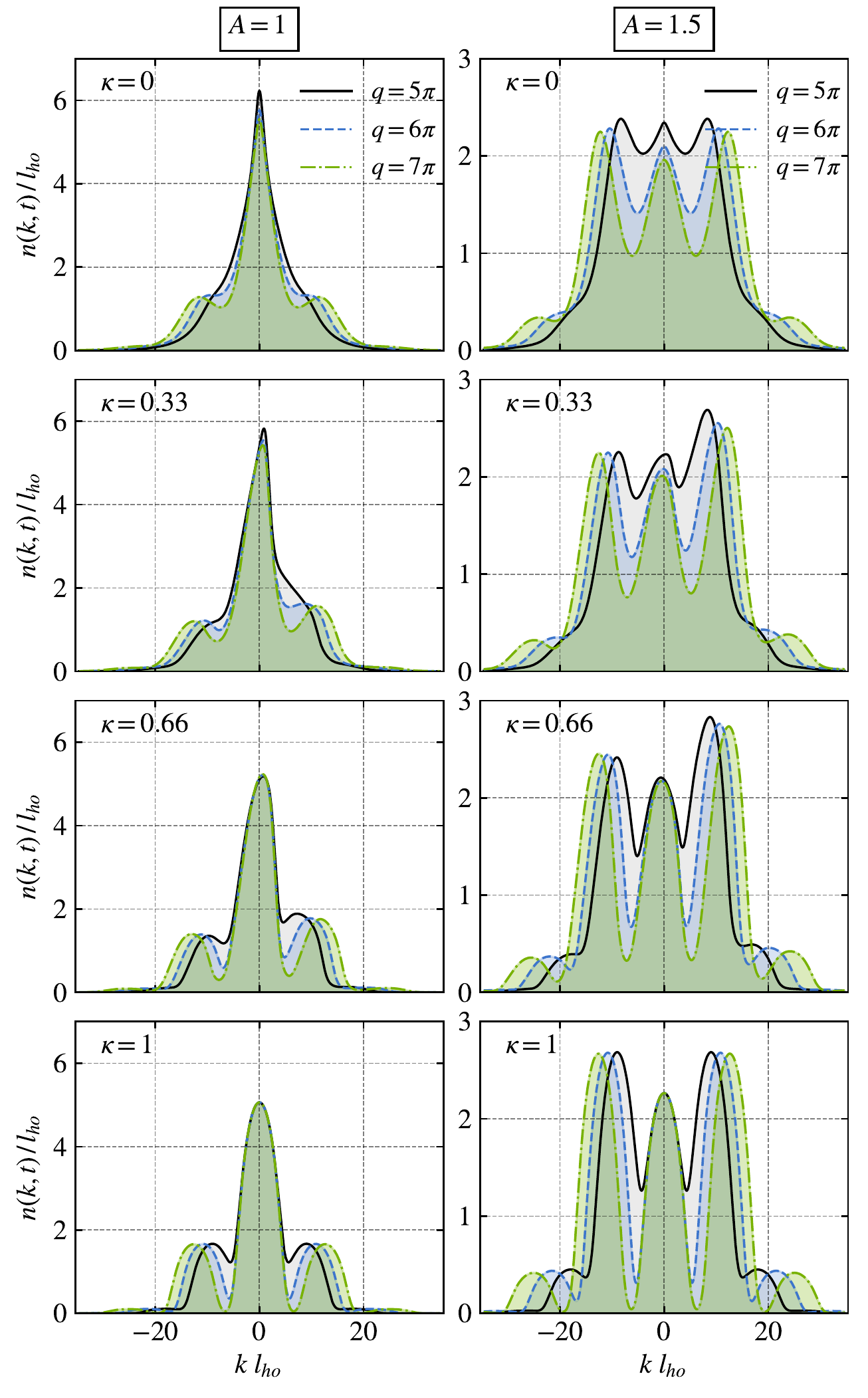}
\caption{Dependence of the relaxed momentum distribution function for $N=10$ particles on the Bragg momentum $q$ and $A$. In
the first column $A=1$ and in the second column $A=1.5$. The initial dimensionless temperature is $\theta_0=0.083$ ($T_0=0.25$), $l_{ho}=0.57$, $\omega=3$,
and $t=\Delta t$  with $\Delta t=\pi/(40 \omega)$. }
\label{depA}
\end{figure}

Using the Jacobi-Anger expansion $e^{z\cos \theta}=I_0(z)+2\sum_{k=1}^\infty I_k(z)\cos\theta$ (9.6.34 of \cite{AS})  with
$I_k(z)=\int_0^\pi e^{z\cos\theta}\cos(k\theta)\, d\theta/\pi$ the  modified Bessel function of the first kind
in the form $e^{-i z \cos\theta}=\sum_{k=-\infty}^\infty I_k(-iz)e^{-i k\theta}$ and the identity (7.374(8) of \cite{GR})
\[
\inti e^{-(u-z)^2}H_j(\alpha u)\, du=\sqrt{\pi}(1-\alpha^2)^{j/2}H_j\left(\frac{\alpha z}{\sqrt{1-\alpha^2}}\right)\, ,
\]
we find \cite{BWENKC}:
\begin{align}
\phi_j(z,t)=&\sum_{n=-\infty}^\infty I_n(-iA)  e^{-i n q \cos (\omega t)\left(z+n \frac{q \sin(\omega t)}{2 m\omega}\right)} \nonumber\\
&\ \ \ \times  \phi_j\left(z+n \frac{ q\sin(\omega t)}{m \omega}\right)  e^{-i\omega\left(j+\frac{1}{2}\right) t} \, .
\end{align}
The wave-functions are periodic in time with period $T=2\pi/\omega$ and at first impression it would seem that also
the real space density and one-particle density matrix have the same periodicity. However, using the fact that
$I_n(-i A)=I_{-n}(-iA)$ it is easy to see that
\be
\phi_j\left(z,t+\frac{\pi}{\omega}\right)=e^{-i\left(j+\frac{1}{2}\right)\pi}\phi_j(z,t)\, ,
\ee
which implies that all the $n$-particle reduced matrices are in fact periodic with period $T=\pi/\omega$.
This is due to the fact that in their definition (\ref{reducedn}) the wave-functions  appear in the
form $\psi_{N,A,\bm{\nu}} \psi_{N,A,\bm{\nu}}^*$ and from the Anyon-Fermi mapping (\ref{AFmap}) we have
\[
\psi_{N,A,\boldsymbol{\nu}}\left(\boldsymbol{z}|\, t+\frac{\pi}{\omega}\right)=A(\boldsymbol{z})B(\boldsymbol{z})
e^{-i\sum_{\nu_j}\left(\nu_j+\frac{1}{2}\right)\pi} \psi_{N,F,\boldsymbol{\nu}}(\boldsymbol{z}|\, t)\, .
\]
Therefore, it will be sufficient to study the dynamics of the real-space density and momentum distribution for $t\in [0,\pi/\omega]$.
We should point out that this periodicity is particular to the TG regime and is no longer valid in the case of finite
coupling.

It will be useful to introduce a dimensionless initial temperature defined by $\theta_0=T_0/N\omega$.
In Fig.~\ref{NCall} we present the dynamics of the momentum distribution and real-space density in the QNC setup for $N=10$
anyons at initial dimensionless temperature $\theta_0=0.083$ with $A=1.5$  and $q=6\pi$ the parameters of the Bragg pulse.
In the first three columns we plot the momentum distributions of impenetrable bosons ($\kappa=0$) and anyons with $\kappa=0.33$ and $\kappa=0.66$
together with the momentum distribution of free fermions ($\kappa=1$). For all values of the statistics parameter the evolution
during a period is similar: immediately after the Bragg pulse the momentum distribution presents local maxima at $\pm q$  and $\pm 2 q$
followed by the well-known oscillations. The distinguishing feature of the anyonic particles is the non-symmetric momentum
distribution which is most visible in the vicinity of $t=p \pi/\omega$ with $p=0,1,2,\cdots$. This is a well-known characteristic
of 1D anyons \cite{HZC1,SC,delC,Patu1} and is a result of the broken space-reversal symmetry  (see (\ref{asymm})
and (\ref{comm})). During the evolution we see that for all values of $\kappa$  the overlap between the momentum distribution of
anyons and free fermions becomes significant and reaches its minimum in the vicinity of  $t=p \pi/\omega$ with $p=0,1,2,\cdots$.
In the bosonic case a similar phenomenon was first described in \cite{MG05} in the context  of breathing oscillations and it was
dubbed (periodic) dynamical fermionization. The dynamics of the real-space density is presented in the last column of Fig.~\ref{NCall}.

It was discovered in \cite{BWENKC} that there are two different time scales in the QNC evolution. In addition to the slow in-trap periodic
behavior presented in Fig.~\ref{NCall} the system exhibits also rapid  relaxation immediately after the pulse which is shown in Fig.~\ref{First}.
The amplitude of this relaxation is largest for the bosonic system and monotonically decreases as we increase the statistics parameter
so it can be said that is directly proportional with the degree of ``interaction"  of the system.
In the case of the fermionic system (which is noninteracting) we see that the momentum distribution is almost unchanged immediately after the pulse.

The dependence of relaxed momentum distribution function on the Bragg momentum $q$  and the parameter $A$ can be found in Fig.~\ref{depA}.
If $A$ is too small the Bragg pulse will not remove the majority of particles from the vicinity of $k=0$ and the periodic
oscillations will not be clearly visible. In the bosonic case for $A=1.5$ the characteristic ghost shape of the relaxed
momentum distribution is clearly visible and becomes more pronounced with increasing Bragg momentum $q$.
Also, as expected, the width of the momentum distribution is  a monotonically increasing function of both  $q$ and $A$.    In the case of anyonic
particles the relaxed distribution is asymmetric with respect to $k=0$.

\begin{figure}
\includegraphics[width=1\linewidth]{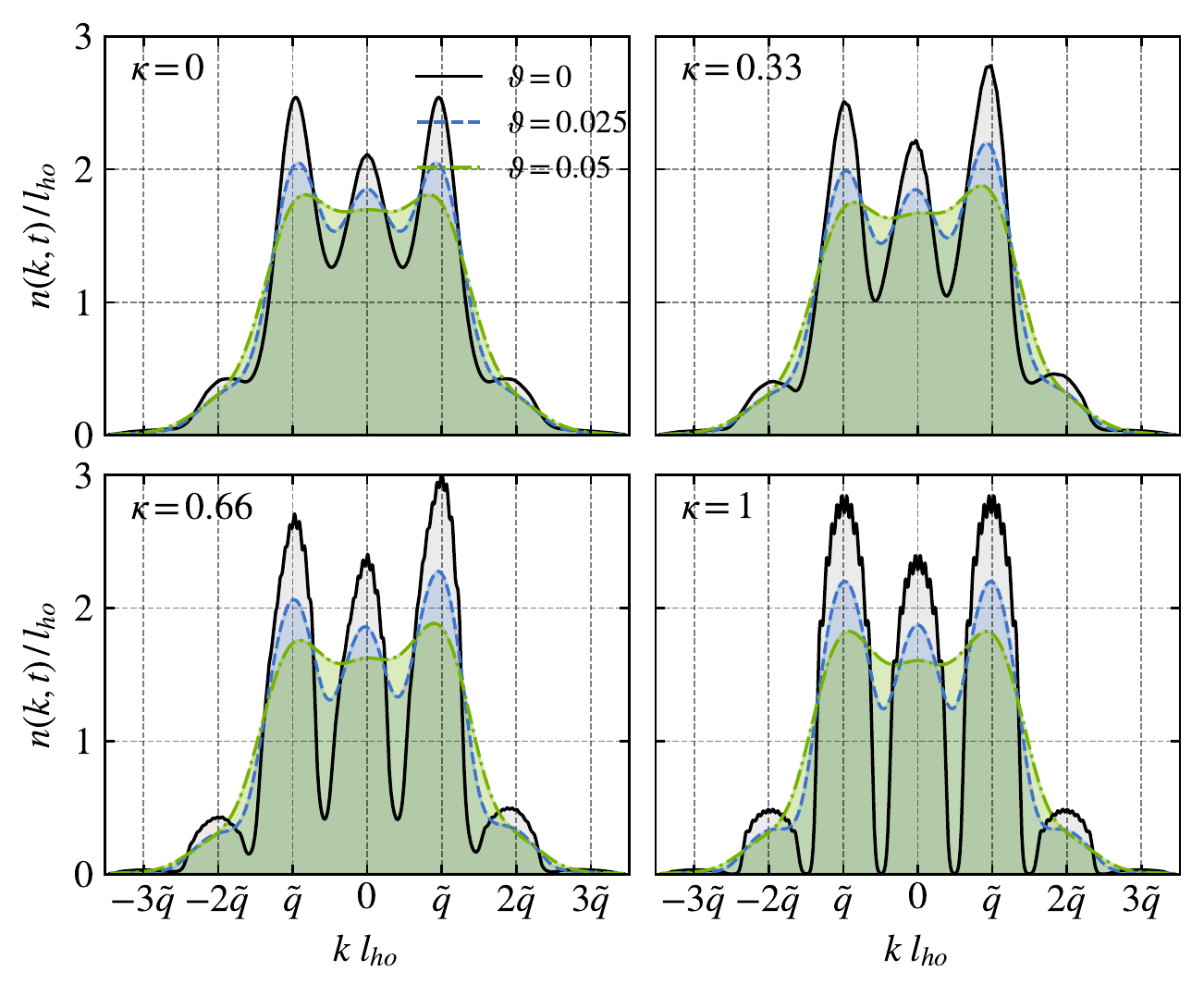}
\caption{Dependence on the initial temperature of the relaxed momentum distribution. Here $N=10$, $A=1.5$, $q=6\pi$, $l_{ho}=0.56$, $\omega=3$, and
$t=2\Delta t$ with $\Delta t=\pi/(40 \omega)$.  }
\label{Tempd}
\end{figure}

The influence of the initial temperature on the relaxed momentum distribution is shown in Fig.~\ref{Tempd}. As the initial temperature
increases the $\pm q$ satellites become less and less pronounced and the observation  of the oscillations becomes harder.

\section{Dynamics in a harmonic trap with time-dependent frequency}\label{s6}

\begin{figure}
\includegraphics[width=1\linewidth]{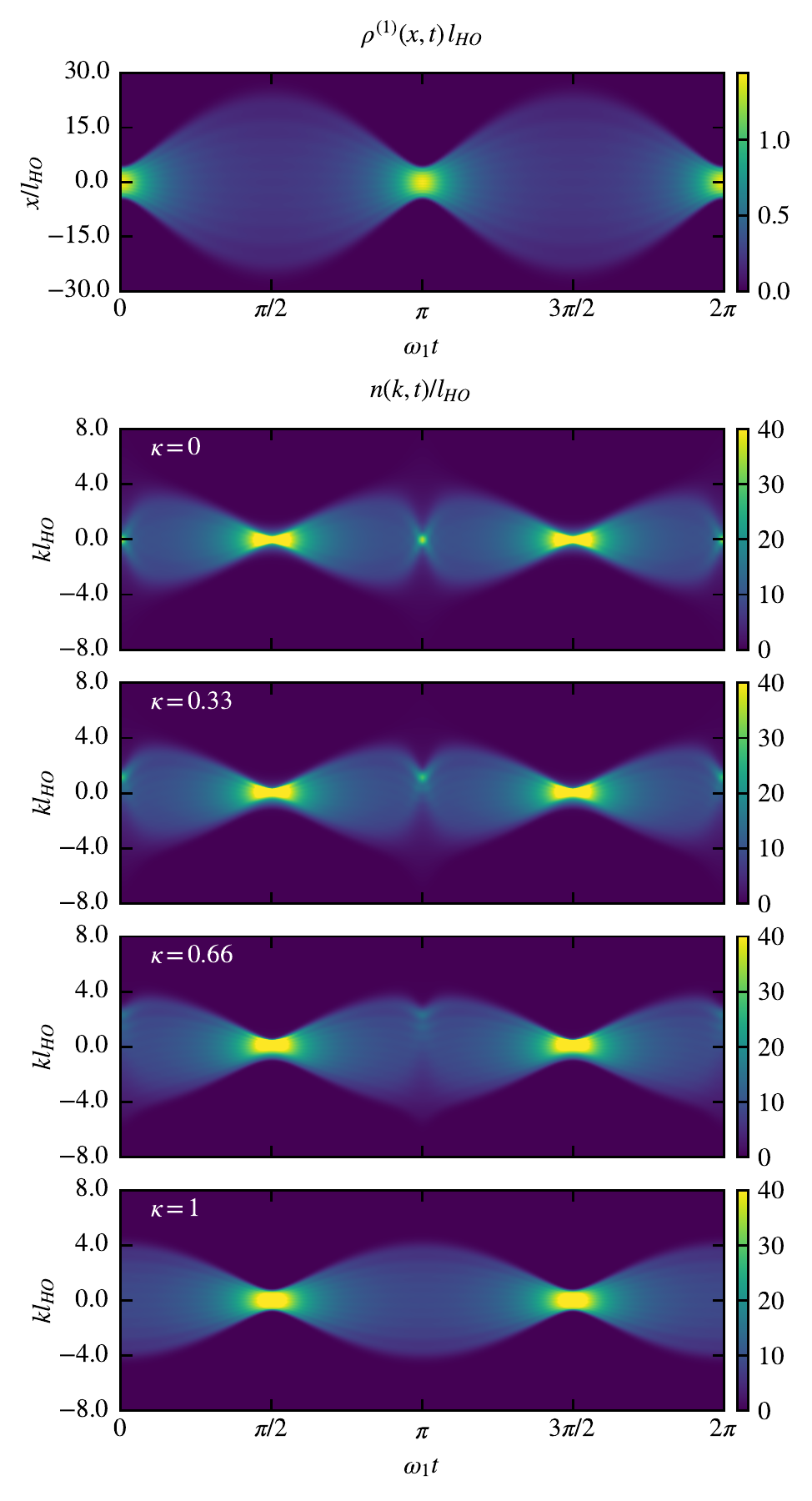}
\caption{Evolution of the real-space density (upper panel) and momentum distribution (lower panels) of an anyonic TG gas of $N=10$
particles  after a confinement quench  of the frequency with $\epsilon=35$ and initial dimensionless temperature
$\theta_0=0.02$ $(T_0=1.2)$ .
}
\label{OscillationsDP}
\end{figure}

\begin{figure*}
\includegraphics[width=1\linewidth]{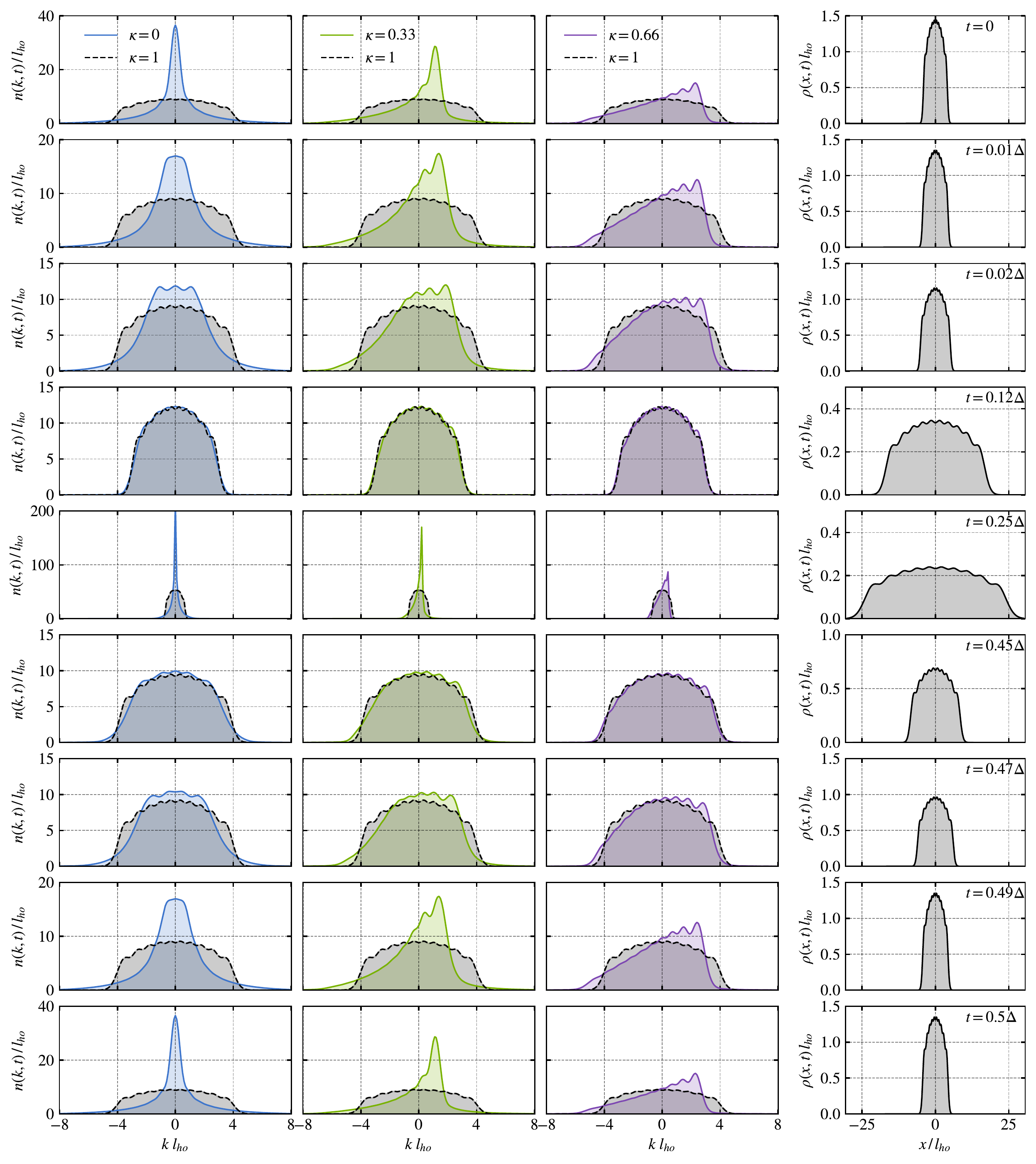}
\caption{Momentum distribution (first three columns) and  real-space density (last column) of an anyonic TG gas after a
confinement quench  of the frequency for selected values of $t$. We use the same parameters as in Fig.~\ref{OscillationsDP}
and $\Delta t= 2\pi /\omega_1$. }
\label{OscillationsDPt}
\end{figure*}

Another experimentally relevant situation that we are going to investigate is the dynamics of a gas in a harmonic
potential with time dependent frequency $V(z,t)= m\omega^2(t) z^2/2$. We are going to denote the frequency of the
potential at $t=0$ by $\omega_0$. At $t=0$  the SP wave-functions are given by (\ref{hermite}) with
$\omega=\omega_0.$ The study of the dynamics in this case is significantly simplified by the fact that the
time-evolution of the wave-functions  is given by the following scaling transformation: (\cite{PP70}, Chap.
VII of \cite{PZ})
\begin{align}\label{scaling}
\phi_j(z,t)=\frac{1}{\sqrt{b(t)}}\phi_j\left(\frac{z}{b(t)},0\right)e^{i\frac{ m x^2}{2}\frac{\dot{b}(t)}{b(t)}-i E_j\tau(t)}\, ,
\end{align}
with $b(t)$ the solution of the second-order differential equation $\ddot{b}+\omega^2(t) b=\omega_0^2/b^3$, also
known as the  Ermakov-Pinney equation,  with initial conditions $b(0)=1, \,  \dot{b}(0)=0$ and $\tau(t)=\int_0^t
dt'/b^2(t')$. Eq.~(\ref{scaling})  represents the unique time-dependent solution of the Schr\"odinger equation
$i\hbar \6 \phi_{j}(z,t)/\6 t=H_{\mbox{\tiny{SP}}}(z,t) \phi_{j}(z,t)$ with $H_{\mbox{\tiny{SP}}}(z,t)=-(\hbar^2/2m)(\6^2/\6 z^2)+V(z,t)$  where
$H_{\mbox{\tiny{SP}}}(z,0)\phi_{j}(z)=\omega_0(j+1/2)\phi_{j}(z)$.

Using the formula for the Slater determinant (\ref{Slater}) and the fact that the Anyon-Fermi mapping is unchanged
by the scaling  transformation the $N$-body anyonic  wave-function satisfies
\begin{align}\label{i11}
\psi_{N,A,\bm{\nu}}(\bm{z}|\, t)&=\frac{1}{b^{N/2}}\psi_{N,A,\bm{\nu}}\left(\bm{z}/b|\, 0\right)e^{-i\sum_jE_{\nu_j}\tau}\nonumber\\
&\ \ \ \ \ \ \ \ \ \ \ \ \ \ \ \ \ \ \ \times e^{i\frac{\dot{b}}{b}\sum_{j}m z_j^2/2}\, .
\end{align}
Inserting (\ref{i11}) in the definition of the one-body density matrix (\ref{reduced}) we find \cite{MG05} (see also \cite{CSC13,RBD19})
\be\label{scalingr}
\rho^{(1)}(x,y|\, t)=\frac{1}{b}\rho^{(1)}\left(\frac{x}{b},\frac{y}{b}\left.\right|\, 0\right)
e^{-i\frac{\dot{b}}{b}\frac{m(x^2-y^2)}{2}}\, .
\ee
The momentum distribution function can also be written as (we perform the change of variables $x'/b,y'/b\rightarrow x,y$)
\be\label{scalingn}
n(k,t)= b\int dx dy\,  \rho^{(1)}(x,y|\, 0)e^{- i b \left[\dot{b}\frac{ m(x^2-y^2)}{2}+k(x-y)\right]}\, .
\ee
The simplification introduced by Eq.~(\ref{scalingr}) is now evident. The study of the dynamics is now reduced
to the calculation of the initial ($t=0$) reduced density matrix at thermal equilibrium supplemented by the
solution of the Ermakov-Pinney equation. In this case the it is preferable to numerically evaluate the initial RDM
using the method of Sec.~\ref{s43} due to the fact that the off-diagonal overlaps can be analytically calculated
\cite{Pic}. Following \cite{AGBKa} we introduce $\xi=z/l_{ho}$  and $\varphi_j(\xi)=\sqrt{l_{ho}}\phi_j(z)$. Then
for $j\ne k$  we have (formula B.16 of \cite{Pic})
\be\label{an1}
\int\varphi_j(\xi)\varphi_k^*(\xi)\, d\xi=e^{-\xi^2} \frac{[H_{j+1}(\xi)H_k(\xi)-H_j(\xi)H_{k+1}(\xi)]}{2(k-j) (2^{j+k}\pi j!k!)^{1/2}}\, .
\ee
For the diagonal elements it seems that a similar formula does not exist but an efficient recursive formula
can be devised as follows \cite{AGBKa}. We define a sequence of functions $\{M_j(\xi)\}_{j=0}^\infty$ with
$M_0(\xi)=0$  and the general term satisfying
\be\label{an2}
M_j(\xi)=\frac{\sqrt{\pi}}{2}\mbox{erf}(\xi)-\frac{1}{2^j j!}\int e^{-\xi^2} H_j^2(\xi)\, d\xi\, ,
\ee
where $\mbox{erf}(\xi)=2\int_0^\xi e^{-t^2}\, dt/\sqrt{\pi}$ is the error function. Using the recurrence relation
for Hermite polynomials $H_{j+1}(\xi)=2\xi H_j(\xi)-H'_j(\xi)$ we find
\be\label{an3}
M_{j+1}(\xi)=M_j(\xi)+\frac{e^{-\xi^2}}{2^{j+1} (j+1)!}H_j(\xi)H_{j+1}(\xi)\, .
\ee
Using (\ref{an1}), (\ref{an2}) and (\ref{an3}) all the elements of the matrix $\boldsymbol{P}$ defined in (\ref{defP})
can be calculated analytically without resorting to time consuming  numerical integration.

We are going to consider the situation when initially the gas is in thermal equilibrium at temperature $T_0$
in the presence of the harmonic potential $V(z)= m\omega_0^2 z^2/2$ and perform an instantaneous change of the
trapping frequency from $\omega_0$ to $\omega_1$ at $t=0$. The quench strength will be parameterized by a
dimensionless parameter $\epsilon=\omega_0^2/\omega_1^2-1$. For this particular quench the Ermakov-Pinney equation
takes the form $\ddot{b}+\omega_1^2 b=\omega_0^2/b^3$ with the solution $b(t)=[1+\epsilon \sin^2(\omega_1 t)]^{1/2}$
which describes periodic oscillations between one and $\omega_0/\omega_1$ with period $T=\pi/\omega_1$.

The breathing-mode dynamics of the anyonic TG gas initiated by the quench of the frequency is presented
in Fig.~\ref{OscillationsDP}. While the real space density displays undamped oscillations with period $T=\pi/\omega_1$
the dynamics of the momentum distribution displays a richer structure. Similar to the case of impenetrable bosons
at zero temperature studied in \cite{MG05} and  the QNC case (see the previous Section) the momentum distribution
of the TG anyonic gas displays alternatively anyonic and fermionic features. This dynamical fermionization occurs
rapidly for $t$ close to zero and $T$ as it can be seen in Fig.~\ref{OscillationsDPt} where we present the momentum
distribution for anyonic gases with  $\kappa=\{0, 0.33, 0.66, 1\}$ at selected values of $t$. The overlap between the
anyonic and fermionic momentum distribution is very large for almost the entire period with the exception of a time
interval around $t=T/2$ where the system recovers the initial anyonic distribution rescaled by a factor $b_{\mbox{\tiny{max}}}>1$.
At zero temperature and for $\kappa=0$ we have $b_{\mbox{\tiny{max}}}=\omega_0/\omega_1$  and $n(k, T/2)= b_{\mbox{\tiny{max}}}
\, n(b_{\mbox{\tiny{max}}} p,0)$ \cite{MG05}.

Another interesting feature of the momentum distribution is that it displays narrowing and broadening
cycles occurring at twice the rate of the real-space density. This phenomenon was first identified in the bosonic case
in \cite{AGBKa}  further studied in \cite{ABGKb} and it can be clearly seen in Fig.~\ref{OscillationsDP}. While the
fermionic system presents narrowing of $n(k,t)$ at $\omega_1t=\pi/2+ \pi l$  with $l=1,2,\cdots$  in the anyonic case
the momentum distribution presents an additional narrowing at $\omega_1 t= \pi l$ when the gas is maximally compressed.
This narrowing is the largest for $\kappa=0$ and as we increase the statistics parameter it decreases in an asymmetric (with respect to $k$)
fashion until it disappears for the fermionic system. In \cite{AGBKa, ABGKb} it was argued that this phenomenon is a collective many-body bounce
effect due to the increased thermodynamic pressure of the maximally compressed gas which acts as a potential barrier.

\section{Conclusions}\label{s7}

In this paper we have derived an exact description of the non-equilibrium dynamics of an anyonic TG gas at finite temperature
generalizing the results of \cite{AGBKa}. The evolution of the anyonic one-particle RDM after a quantum quench from an
initial thermal state is given by the Fredholm minor of an integral operator with the kernel being the thermal RDM of free fermions.
The statistics parameter enters  in the constant in front of the integral operator. We have argued that when the overlaps of the
evolved wave-functions cannot be calculated analytically the most efficient numerical evaluation of the RDM is based on Nystr\"{o}m's
method of solving Fredholm integral equations of the second kind.  We have investigated the non-equilibrium dynamics of the
anyonic TG gas in two experimental relevant scenarios: the QNC and the breathing oscillations initiated
by a sudden quench of the trapping frequency. A natural extension of our work would be to investigate the finite temperature
dynamics of the bipartite entanglement as it was already noticed in \cite{CDM} that Nystr\"{o}m's method is more efficient in
computing the entanglement entropy in the situations in which the overlap matrix cannot be computed analytically. Other promising avenues
of research are the investigation of the anyonic spectral function \cite{LGPM20}, entanglement revival \cite{MAC20} or the non-equilibrium quantum thermodynamics \cite{ASK20}.  This will
be deferred to future publications.

\acknowledgments

Financial support  from the LAPLAS 6 program of the Romanian National Authority for Scientific Research (CNCS-UEFISCDI) is gratefully acknowledged.

\appendix

\section{Fredholm minors}\label{a1}

In this Appendix we present some basic definitions and useful formulas involving Fredholm minors which were
used in the main text. Consider a Fredholm integral equation of the second kind (for more information
on Fredholm integral equation see Chap. II of \cite{Pogorz} or Chap. I of \cite{Smirn4})
\be\label{ai1}
g(\lambda)=f(\lambda)+\gamma\int_{\Omega} \textsf{K}(\lambda,\mu)g(\mu)\, d\mu\, ,
\ee
where $f(\lambda)$ is a continuous complex function defined on the bounded set $\Omega$ and the
kernel $\textsf{K}(\lambda,\mu)$ is a continuous complex function of $\lambda$  and $\mu$
on $\Omega\times \Omega$. The set $\Omega$  may be a bounded interval or the reunion of a
finite number of such intervals. The $n$-th Fredholm minor of the integral operator $\hat{\textsf{K}}$
is given by the series
\begin{widetext}
\be\label{defm}
D_n \left(\left.\begin{array}{ccc}
              \lambda_1& \cdots&\lambda_n\\
              \mu_1& \cdots& \mu_n
         \end{array} \right| \gamma
\right)=
\textsf{K} \left(\begin{array}{ccc}
              \lambda_1& \cdots& \lambda_n\\
              \mu_1& \cdots& \mu_n
           \end{array}
\right)+
\sum_{p=1}^\infty \frac{(-\gamma)^p}{p!}\int_\Omega\cdots\int_{\Omega}
\textsf{K}\left(\begin{array}{cccccc}
              \lambda_1& \cdots & \lambda_n & \nu_1 &\cdots &\nu_p\\
              \mu_1& \cdots & \mu_ n& \nu_1 &\cdots &\nu_p
          \end{array}
\right) \, d\nu_1 \cdots d\nu_p\, ,
\ee
where we have introduced the notation
\be\label{ai2}
\textsf{K} \left(\begin{array}{ccc}
              \lambda_1& \cdots& \lambda_n\\
              \mu_1& \cdots& \mu_n
           \end{array}
           \right) =
\left|
\begin{array}{cccc}
\textsf{K}(\lambda_1,\mu_1)& \textsf{K}(\lambda_1,\mu_2)& \cdots & \textsf{K}(\lambda_1,\mu_n)\\
\textsf{K}(\lambda_2,\mu_1)& \textsf{K}(\lambda_2,\mu_2)& \cdots & \textsf{K}(\lambda_2,\mu_n)\\
\vdots &\vdots&\ddots&\vdots\\
\textsf{K}(\lambda_n,\mu_1)& \textsf{K}(\lambda_n,\mu_2)& \cdots & \textsf{K}(\lambda_n,\mu_n)\\
\end{array}
\right|\, .
\ee
The series (\ref{defm}) converges for all values of the parameter $\gamma$ and $D_n$ is antisymmetric in both $\lambda_i$'s
and $\mu_i$'s by construction. Particular cases of the series which will play an important role in our analysis are the
$n=0$ case known as the Fredholm determinant
\be\label{ai3}
D(\gamma)\equiv \mbox{\textsf{det}}(1-\gamma \hat{\textsf{K}})=1+
\sum_{p=1}^\infty \frac{(-\gamma)^p}{p!}\int_\Omega\cdots\int_{\Omega}
\textsf{K}\left(\begin{array}{ccc}
               \nu_1 &\cdots &\nu_p\\
               \nu_1 &\cdots &\nu_p
          \end{array}
\right) \, d\nu_1 \cdots d\nu_p\, ,
\ee
and the first minor $(n=1)$
\be\label{ai4}
D \left(\left.\begin{array}{c}
              \lambda\\
              \mu
         \end{array} \right| \gamma
\right)=
\textsf{K}(\lambda,\mu)
+
\sum_{p=1}^\infty \frac{(-\gamma)^p}{p!}\int_\Omega\cdots\int_{\Omega}
\textsf{K}\left(\begin{array}{cccc}
              \lambda & \nu_1 &\cdots &\nu_p\\
              \mu& \nu_1 &\cdots &\nu_p
          \end{array}
\right) \, d\nu_1 \cdots d\nu_p\, .
\ee
\end{widetext}
If $D(\gamma)\ne 0$ a solution of the integral equation (\ref{ai1}) is given by
\be
g(\lambda)=f(\lambda)+\gamma\int_{\Omega} \textsf{R}(\lambda,\mu) f(\mu)\, d\mu\, ,
\ee
with the function $\textsf{R}(\lambda,\mu)$ called the resolvent kernel which
satisfies
\be\label{ai5}
\textsf{R}(\lambda,\mu)=\textsf{K}(\lambda,\mu)+\gamma\int_{\Omega}\textsf{K}(\lambda,\nu)\textsf{R}(\nu,\mu)\, d\nu\, ,
\ee
and is given by
\be\label{ai6}
\textsf{R}(\lambda,\mu)=D \left(\left.\begin{array}{c}
              \lambda\\
              \mu
         \end{array} \right| \gamma
\right)/ \mbox{\textsf{det}}(1-\gamma \hat{\textsf{K}})\, .
\ee
Formula (\ref{ai6}) is a particular case of a more general identity first proved by Hurwitz \cite{Hurw14},
and possibly rediscovered many times \cite{Fein04}, which relates the $n$-th minor to the resolvent kernel
\be\label{ai7}
D_n \left(\left.\begin{array}{ccc}
              \lambda_1& \cdots& \lambda_n\\
              \mu_1& \cdots& \mu_n
         \end{array} \right| \gamma
\right)=
 \mbox{\textsf{det}}(1-\gamma \hat{\textsf{K}})
\textsf{R} \left(\begin{array}{ccc}
              \lambda_1& \cdots& \lambda_n\\
              \mu_1& \cdots& \mu_n
         \end{array}
\right)\, .
\ee

\section{Reduced density matrices for free fermions}\label{a2}

Here we compute the $n$-body RDMs for free fermions in a
time-dependent confining potential following  Lenard \cite{Len66}.  Using the symmetry of the fermionic
wave-functions we see that an alternative definition for the fermionic RDMs
is given by
\begin{align}\label{reducednf}
\rho^{(n)}_F(\boldsymbol{x},\boldsymbol{y}|\, t)=&\sum_{N=n}^\infty\sum_{\boldsymbol{\nu}}p(N,\boldsymbol{\nu}) \frac{N!}{(N-n)!}\int dz_{n+1}\cdots dz_{N}\nonumber\\
& \times\psi_{N,F,\boldsymbol{\nu}}(\boldsymbol{x},\boldsymbol{\tilde{z}}|\,t) \psi_{N,F,\boldsymbol{\nu}}^*(\boldsymbol{y},\boldsymbol{\tilde{z}}|\,t)\, ,
\end{align}
with $\boldsymbol{x}=(x_1,\cdots,x_n)\, ,\boldsymbol{y}=(y_1,\cdots,y_n)$ and $\boldsymbol{\tilde{z}}=(z_{n+1},\cdots,z_{N})$.
The wave-functions are given by the Slater determinants (\ref{Slater}) and we are going to consider the grand-canonical ensemble with
$p(N,\boldsymbol{\nu})=e^{-(E_{N,\boldsymbol{\nu}} -\mu N)/k_B T_0}/\mathcal{Z}$.
\begin{widetext}
As a starting point we will derive a preliminary result. Consider
\be
G=\sum_{\boldsymbol{\nu}}e^{-(E_{N,\boldsymbol{\nu}} -\mu N)/k_B T_0}
\psi_{N,F,\boldsymbol{\nu}}(x_1,\cdots,x_N|\, t)\psi^*_{N,F,\boldsymbol{\nu}}(y_1,\cdots,y_N|\, t)\, .
\ee
The summand is symmetric in $\nu_i$'s and vanishes when two of them are equal which means that
$G$ can written as
\begin{align}
G=\frac{1}{N!}\sum_{\nu_1}\cdots\sum_{\nu_N} e^{-(\sum_i E_{\nu_i} -\mu N)/k_B T_0}
\psi_{N,F,\boldsymbol{\nu}}(x_1,\cdots,x_N|\, t)\psi^*_{N,F,\boldsymbol{\nu}}(y_1,\cdots,y_N|\, t)\, .
\end{align}
Using the sum over the permutations form of the determinant we find
\begin{align}
G=&\frac{1}{(N!)^2}\sum_{\nu_1}\cdots\sum_{\nu_N} \left(\prod_{j=1}^N e^{-(E_{\nu_j} -\mu N)/k_B T_0}\right)
\left(\sum_{P\in S_N}(-1)^P \prod_{j=1}^N \phi_{\nu_j}(x_{P(j)},t)\right)
\left(\sum_{Q\in S_N}(-1)^Q \prod_{j=1}^N \phi_{\nu_j}^*(y_{Q(j)},t)\right)
\, ,\nonumber\\
=&\frac{1}{(N!)^2}\sum_{\nu_1}\cdots\sum_{\nu_N} \left(\prod_{j=1}^N e^{-(E_{\nu_j} -\mu N)/k_B T_0}\right)
\sum_{P\in S_N} \sum_{R\in S_N}(-1)^R \prod_{j=1}^N \phi_{\nu_j}(x_{P(j)},t)\phi_{\nu_j}^*(y_{RP(j)},t)\, ,\nonumber\\
=& \frac{1}{(N!)^2}  \sum_{P\in S_N} \sum_{R\in S_N}(-1)^R \prod_{j=1}^N \sum_{\nu_j} e^{-(E_{\nu_j} -\mu N)/k_B T_0}
\phi_{\nu_j}(x_{P(j)},t)\phi_{\nu_j}^*(y_{RP(j)},t)\, ,\nonumber\\
=&\frac{1}{N!}\,\textsf{f}\left(\begin{array}{ccc}
                  x_1&\cdots&x_N\\
                  y_1&\cdots&y_N
                  \end{array}
                  ;t
\right)\, ,\ \  \  \textsf{f}(x,y|\, t)= \sum_{\nu}  e^{-(E_{\nu} -\mu N)/k_B T_0} \phi_{\nu}(x,t)\phi_{\nu}^*(y,t)\, .
\end{align}
where $S_N$ is the group of permutations of $N$ elements, $(-1)^P$ is the signature of the permutation $P$,  and in the second
line we have used the fact that every permutation $Q$ can be written as $Q=RP$ with  signature $(-1)^{R+P}$. In a similar fashion
we can compute the grand-canonical partition function as
\begin{align}\label{ai8}
\mathcal{Z}=&\sum_{N=0}^\infty \sum_{\boldsymbol{\nu}} e^{-(E_{N,\boldsymbol{\nu}} -\mu N)/k_B T_0}
=\sum_{N=0}^\infty \sum_{\boldsymbol{\nu}} e^{-(E_{N,\boldsymbol{\nu}} -\mu N)/k_B T_0}
\int dz_1\cdots\int dz_N\, |\psi_{N,F,\boldsymbol{\nu}}(z_1,\cdots,z_N)|^2\, ,\nonumber\\
=&\sum_{N=0}^\infty\frac{1}{N!}\int dz_1\cdots\int dz_N\, \textsf{f}\left(\begin{array}{ccc}
                  x_1&\cdots&x_N\\
                  y_1&\cdots&y_N
                  \end{array}
                  ;t
\right) =\mbox{\textsf{det}} (1+\hat{\textsf{f}})\, ,
\end{align}
which shows that the partition function can be expressed as the Fredholm determinant of an integral operator with
kernel $\textsf{f}(x,y|\, t)$. Inserting $p(N,\nu)=e^{-(E_{N,\nu}-\mu N)/T}/Z$ in Eq.~(\ref{reducednf}), we find $\rho_F^{(n)}(x,y|t)=H/Z$ where
\begin{align}\label{ai9}
H=&\sum_{N=n}^\infty \sum_{\boldsymbol{\nu}} e^{-(E_{N,\boldsymbol{\nu}} -\mu N)/k_B T_0}\frac{N!}{(N-n)!}
\int dz_{n+1}\cdots\int dz_N
\psi_{N,F,\boldsymbol{\nu}}(\boldsymbol{x},\boldsymbol{\tilde{z}},|\,t) \psi_{N,F,\boldsymbol{\nu}}^*(,\boldsymbol{y},\boldsymbol{\tilde{z}},|\,t)\, ,\nonumber\\
=&\sum_{N=n}^\infty\frac{1}{(N-n)!}\int dz_{n+1}\cdots\int dz_N\, \textsf{f}\left(\begin{array}{cccccc}
                  x_1&\cdots&x_n& z_{n+1}&\cdots& z_N\\
                  y_1&\cdots&y_n& z_{n+1}&\cdots& z_N
                  \end{array}
                  ;t
\right)\, ,\nonumber\\
=&   D_{n}\left.\left(\begin{array}{ccc}
                  x_1&\cdots&x_n\\
                  y_1&\cdots&y_n
                  \end{array}
                  \right|-1
\right)\, ,
\end{align}
\end{widetext}
which shows that $H$ is the $n$-th Fredholm minor of the integral operator with kernel $\textsf{f}(x,y|\, t)$ and
$\gamma=-1$ (see \ref{defm}). Collecting (\ref{ai8}) and (\ref{ai9}) and plugging in the definition (\ref{reducednf})
we find
\be\label{ai10}
\rho^{(n)}_F(\boldsymbol{x},\boldsymbol{y}|\, t)= D_{n}\left.\left(\begin{array}{ccc}
                  x_1&\cdots&x_n\\
                  y_1&\cdots&y_n
                  \end{array}
                  \right|-1
\right)/\mbox{\textsf{det}} (1+\hat{\textsf{f}})\, .
\ee
We can simplify further this expression by considering the resolvent kernel associated to the integral
operator $1+\hat{\textsf{f}}$ which will be denoted by  $F(\lambda,\mu|\,t)$  and satisfies the
integral equation
\be\label{ai11}
F(\lambda,\mu|\,t)+\int \textsf{f}(\lambda, \nu|\, t)F(\nu,\mu|\,t)\, d\nu=\textsf{f}(\lambda, \mu|\, t)\, .
\ee
Expanding $F(\lambda,\mu|\,t)$ in the orthonormal system $\phi_i(\lambda,t) \phi_j^*(\mu,t)$ we obtain
\be\label{ai12}
F(\lambda,\mu|\,t)=\sum_{i=0}^\infty\frac{1}{1+ e^{(E_i-\mu)/k_B T_0}}\phi_i(\lambda,t) \phi_i^*(\mu,t)\, ,
\ee
which shows that $F(\lambda,\mu|\,t)$ is in fact $\rho_F^{(1)}(x,y|\, t)$ the one-body RDM of free fermions.
From (\ref{ai10}) and Hurwitz's identity (\ref{ai7}) we obtain the final expression for the
$n$-body reduced density matrices of free fermions
\be\label{redff}
\rho^{(n)}_F(\boldsymbol{x},\boldsymbol{y}|\, t)=
\rho^{(1)}_F\left(\begin{array}{ccc}
                  x_1&\cdots&x_n\\
                  y_1&\cdots&y_n
                  \end{array}
                 ;t \right)\, .
\ee
with $\rho_F^{(1)}(\lambda,\mu|\,t)=F(\lambda,\mu|\,t)$ defined in (\ref{ai12}). For a recent utilization of (\ref{ai10}) and Lenard's formula for $n=2$ in
the context of the full counting statistics of the bosonic Tonks-Girardeau gas see \cite{DCVA20}.

\section{Two theorems on Gram determinants}\label{a3}

Here we state and prove two theorems on Gram determinants which are useful in deriving the
identity (\ref{tdet}) which establishes the equivalence in the truncated basis of the Fredholm
determinant with the determinant of the overlaps.

\begin{thm}\label{thm1}
Let $\boldsymbol{a}_1, \cdots,\boldsymbol{a}_n$ be $n$ linearly independent vectors in an $M$-dimensional space
with $\boldsymbol{a}_i=(a_{i1},\cdots,a_{iM})$  and the scalar product $\boldsymbol{a}_i\cdot \boldsymbol{a}_j=\sum_{k=1}^M a_{ik} a_{jk}^*$
(the star denotes complex conjugation). Then the Gram determinant of the vectors  $\boldsymbol{a}_1, \cdots,\boldsymbol{a}_n$
defined by
\be
\Gamma=\left|
\begin{array}{cccc}
        \boldsymbol{a}_1^2    &      \boldsymbol{a}_1\cdot\boldsymbol{a}_2  &   \cdots     &\boldsymbol{a}_1\cdot\boldsymbol{a}_n\\
        \boldsymbol{a}_2\cdot\boldsymbol{a}_1   &   \boldsymbol{a}_2^2      &      \cdots     &    \boldsymbol{a}_2\cdot\boldsymbol{a}_n\\
        \vdots & \vdots & \ddots  & \vdots\\
        \boldsymbol{a}_n\cdot \boldsymbol{a}_1  & \boldsymbol{a}_n\cdot \boldsymbol{a}_2   &\cdots    & \boldsymbol{a}_n^2
        \end{array}
\right|
\ee
can be expressed as
\begin{subequations}
\begin{align}
\Gamma=&\sum_{1\le k_1<\cdots<k_n\le M}
\left|
\begin{array}{cccc}
        a_{1k_1} & a_{1k_2} &\cdots &a_{1k_n}\\
        a_{2k_1} & a_{2k_2} &\cdots &a_{2k_n}\\
        \vdots & \vdots & \ddots  & \vdots\\
         a_{nk_1} & a_{nk_2} &\cdots &a_{nk_n}\\
        \end{array}
\right|^2\, ,\label{gram1}\\
=&\frac{1}{n!}\sum_{k_1=1}^M\cdots\sum_{k_n=1}^M
\left|
\begin{array}{cccc}
        a_{1k_1} & a_{1k_2} &\cdots &a_{1k_n}\\
        a_{2k_1} & a_{2k_2} &\cdots &a_{2k_n}\\
        \vdots & \vdots & \ddots  & \vdots\\
         a_{nk_1} & a_{nk_2} &\cdots &a_{nk_n}\\
        \end{array}
\right|^2\, ,\label{gram2}
\end{align}
\end{subequations}
where $|A|^2=|A||A^*|$ for any matrix $A$.
\end{thm}

\textit{Proof.} Using the linearity of the determinant with respect to columns we have
\begin{align}
\Gamma=&\left|
\begin{array}{cccc}
        \sum_{k_1=1}^M a_{1k_1} a_{1k_1}^*    &      \boldsymbol{a}_1\cdot\boldsymbol{a}_2  &   \cdots     &\boldsymbol{a}_1\cdot\boldsymbol{a}_n\\
        \sum_{k_1=1}^M a_{2k_1} a_{1k_1}^*    &   \boldsymbol{a}_2^2      &      \cdots     &    \boldsymbol{a}_2\cdot\boldsymbol{a}_n\\
        \vdots & \vdots & \ddots  & \vdots\\
        \sum_{k_1=1}^M a_{nk_1} a_{1k_1}^* & \boldsymbol{a}_n\cdot \boldsymbol{a}_2   &\cdots    & \boldsymbol{a}_n^2
        \end{array}
\right|\, , \nonumber\\
=&\sum_{k_1=1}^M a_{1k_1}^*\left|
\begin{array}{cccc}
       a_{1k_1}    &      \boldsymbol{a}_1\cdot\boldsymbol{a}_2  &   \cdots     &\boldsymbol{a}_1\cdot\boldsymbol{a}_n\\
       a_{2k_1}    &   \boldsymbol{a}_2^2      &      \cdots     &    \boldsymbol{a}_2\cdot\boldsymbol{a}_n\\
        \vdots & \vdots & \ddots  & \vdots\\
       a_{nk_1} & \boldsymbol{a}_n\cdot \boldsymbol{a}_2   &\cdots    & \boldsymbol{a}_n^2
        \end{array}
\right|\, ,\nonumber\\
=& \sum_{k_1=1}^M\cdots\sum_{k_n=1}^M a_{1k_1}^* a_{2k_2}^*\cdots a_{nk_n}^*\nonumber\\
&\ \ \ \ \ \ \ \ \ \ \ \ \ \  \times
\left|
\begin{array}{cccc}
        a_{1k_1} & a_{1k_2} &\cdots &a_{1k_n}\\
        a_{2k_1} & a_{2k_2} &\cdots &a_{2k_n}\\
        \vdots & \vdots & \ddots  & \vdots\\
         a_{nk_1} & a_{nk_2} &\cdots &a_{nk_n}\\
        \end{array}
\right|
\end{align}
From the last relation we cans see that when two $k$'s are equal the summand vanishes which means that
we have
\begin{align}
\Gamma=& \sum_{1\le k_1\cdots k_n\le M} \sum_{P\in S_N} a_{1k_{P(1)}}^*a_{2k_{P(2)}}^*\cdots a_{nk_{P(n)}}^*\nonumber\\
&\ \ \ \ \ \ \ \ \ \ \ \ \ \  \times
\left|
\begin{array}{cccc}
        a_{1k_{P(1)}} & a_{1k_{P(2)}} &\cdots &a_{1k_{P(n)}}\\
        a_{2k_{P(1)}} & a_{2k_{P(2)}} &\cdots &a_{2k_{P(n)}}\\
        \vdots & \vdots & \ddots  & \vdots\\
         a_{nk_{P(1)}} & a_{nk_{P(2)}} &\cdots &a_{nk_{P(n)}}\\
        \end{array}
\right|\, \nonumber\\
=& \sum_{1\le k_1\cdots k_n\le M} \left(\sum_{P\in S_N} (-1)^P a_{1k_{P(1)}}^* a_{2k_{P(2)}}^*\cdots a_{nk_{P(n)}}^*\right)\nonumber\\
&\ \ \ \ \ \ \ \ \ \ \ \ \ \  \times
\left|
\begin{array}{cccc}
        a_{1k_1} & a_{1k_2} &\cdots &a_{1k_n}\\
        a_{2k_1} & a_{2k_2} &\cdots &a_{2k_n}\\
        \vdots & \vdots & \ddots  & \vdots\\
         a_{nk_1} & a_{nk_2} &\cdots &a_{nk_n}\\
        \end{array}
\right|\, , \nonumber\\
=&\sum_{1\le k_1<\cdots<k_n\le M}
\left|
\begin{array}{cccc}
        a_{1k_1} & a_{1k_2} &\cdots &a_{1k_n}\\
        a_{2k_1} & a_{2k_2} &\cdots &a_{2k_n}\\
        \vdots & \vdots & \ddots  & \vdots\\
         a_{nk_1} & a_{nk_2} &\cdots &a_{nk_n}\\
        \end{array}
\right|^2\, ,
\end{align}
which proves (\ref{gram1}). Then (\ref{gram2}) follows from the fact that the
square modulus of the determinant is a symmetric function in $k$'s and vanishes
when two of them are equal.

\begin{thm}\label{thm2}

Let $\phi_1(x),\cdots,\phi_n(x)$ linear independent functions defined on the interval $[a,b]$. Then
\begin{align}\label{gram3}
\Gamma=&\left|\int_a^b\phi_i(x)\phi_k^*(x)\, dx\right|\, ,\\
=&\frac{1}{n!}\int_a^b dx_1\cdots\int_a^b dx_n
&\left|
\begin{array}{cccc}
\phi_1(x_1) & \phi_1(x_2) &\cdots& \phi_1(x_n)\\
\phi_2(x_1) & \phi_2(x_2) &\cdots& \phi_2(x_n)\\
\vdots & \vdots & \ddots  & \vdots\\
\phi_n(x_1) & \phi_n(x_2) &\cdots& \phi_n(x_n)\\
\end{array}
\right|^2\, .\nonumber
\end{align}
\end{thm}

\textit{Proof.} The Gram determinant can also be written as
\begin{align*}
\Gamma=&\int_a^b dx_1 \cdots \int_a^b dx_n\nonumber\\
&\times\left|
\begin{array}{cccc}
\phi_1(x_1)\phi_1^*(x_1) & \phi_1(x_2)\phi_2^*(x_2) & \cdots & \phi_1(x_n)\phi_n^*(x_n)\\
\phi_2(x_1)\phi_1^*(x_1) & \phi_2(x_2)\phi_2^*(x_2) & \cdots & \phi2(x_n)\phi_n^*(x_n)\\
\vdots & \vdots & \ddots  & \vdots\\
\phi_n(x_1)\phi_1^*(x_1) & \phi_n(x_2)\phi_2^*(x_2) & \cdots & \phi_n(x_n)\phi_n^*(x_n)\\
\end{array}
\right|\, ,\nonumber\\
=&\int_a^b dx_1 \cdots \int_a^b dx_n\, \phi_1^*(x_1)\cdots\phi_n^*(x_n)\nonumber\\
&\ \ \ \ \ \ \ \ \ \ \ \ \ \ \ \ \times\left|
\begin{array}{cccc}
\phi_1(x_1) & \phi_1(x_2) & \cdots & \phi_1(x_n))\\
\phi_2(x_1) & \phi_2(x_2) & \cdots & \phi_2(x_n))\\
\vdots & \vdots & \ddots  & \vdots\\
\phi_n(x_1) & \phi_n(x_2) & \cdots & \phi_n(x_n)\\
\end{array}
\right|
\end{align*}
due to the fact that in the first determinant the integration variable $x_j$ appears only
in the $j$-th column. In the last relation we can see that the right hand side is unchanged
if we permute the integration variable. Then we can write
 \begin{align*}
\Gamma =&\frac{1}{n!}\int_a^b dx_1 \cdots \int_a^b dx_n\, \sum_{P\in S_N}\phi_1^*(x_{P(1)})\cdots\phi_n^*(x_{P(n)})\nonumber\\
&\ \ \ \ \ \ \ \ \ \times\left|
\begin{array}{cccc}
\phi_1(x_{P(1)}) & \phi_1(x_{P(2)}) & \cdots & \phi_1(x_{P(n)}))\\
\phi_2(x_{P(1)}) & \phi_2(x_{P(2)}) & \cdots & \phi_2(x_{P(n)}))\\
\vdots & \vdots & \ddots  & \vdots\\
\phi_n(x_{P(1)}) & \phi_n(x_{P(2)}) & \cdots & \phi_n(x_{P(n)})\\
\end{array}
\right|\\
=&\frac{1}{n!}\int_a^b dx_1 \cdots \int_a^b dx_n\, \\
&\ \ \ \ \ \ \ \ \ \ \ \ \ \times\left(\sum_{P\in S_N}(-1)^P\phi_1^*(x_{P(1)})\cdots\phi_n^*(x_{P(n)})\right)\nonumber\\
&\ \ \ \ \ \ \ \ \ \ \ \ \ \times\left|
\begin{array}{cccc}
\phi_1(x_1) & \phi_1(x_2) & \cdots & \phi_1(x_n))\\
\phi_2(x_1) & \phi_2(x_2) & \cdots & \phi_2(x_n))\\
\vdots & \vdots & \ddots  & \vdots\\
\phi_n(x_1) & \phi_n(x_2) & \cdots & \phi_n(x_n)\\
\end{array}
\right|
\end{align*}
which proves our theorem. Thm. \ref{thm2} is also known as Andr\'{e}ief's integration formula \cite{Forr18}.

\section{Comparison of the two numerical methods}\label{a4}

\begin{figure*}
\includegraphics[width=1\linewidth]{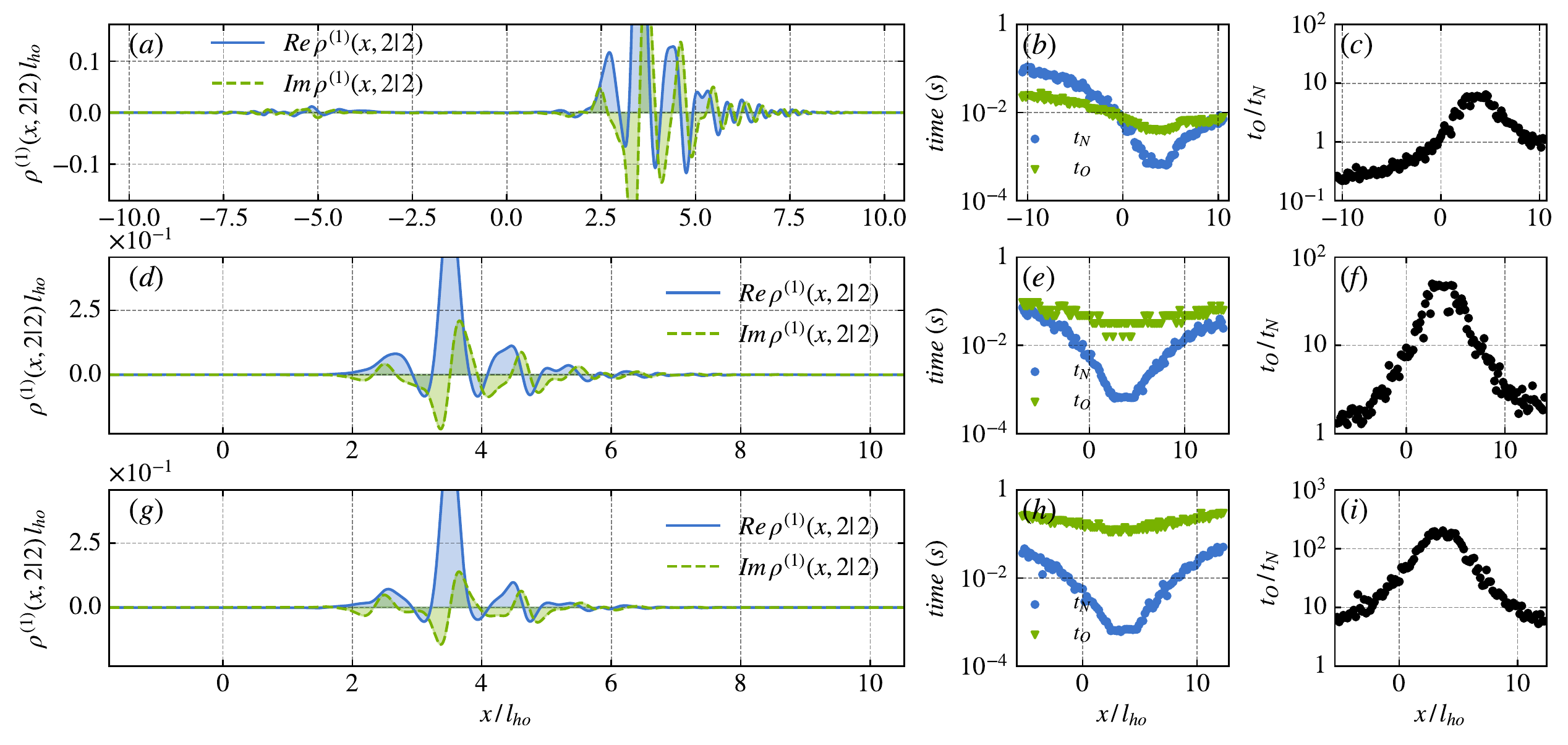}
\caption{First column: Real and imaginary part of  $\rho^{(1)}(x,y|\, t)$ for $N=10$ bosons ($\kappa=0$ and fixed values of $y$ and $t$)
at different temperatures in the quantum Newton's cradle setting ($l_{ho}=0.56, \omega=3, q=4\pi, A=1.5$). Here $y=2,$  $t=2\Delta t, \Delta t=\pi/ 40 \omega$, $\theta_0=0$ (a), $\theta_0=0.011$ (d), and $\theta_0=0.025$ (g).
Middle column: Evaluation time (logarithmic scale) of the RDM using  Nystr\"{o}m's method $t_N$ (disks) and overlaps method $t_O$ (triangles) for the
data in first column. Last column: Plots of the ratios  $t_O/t_N$  from the middle column.  }
\label{figt2}
\end{figure*}

\begin{figure*}
\includegraphics[width=1\linewidth]{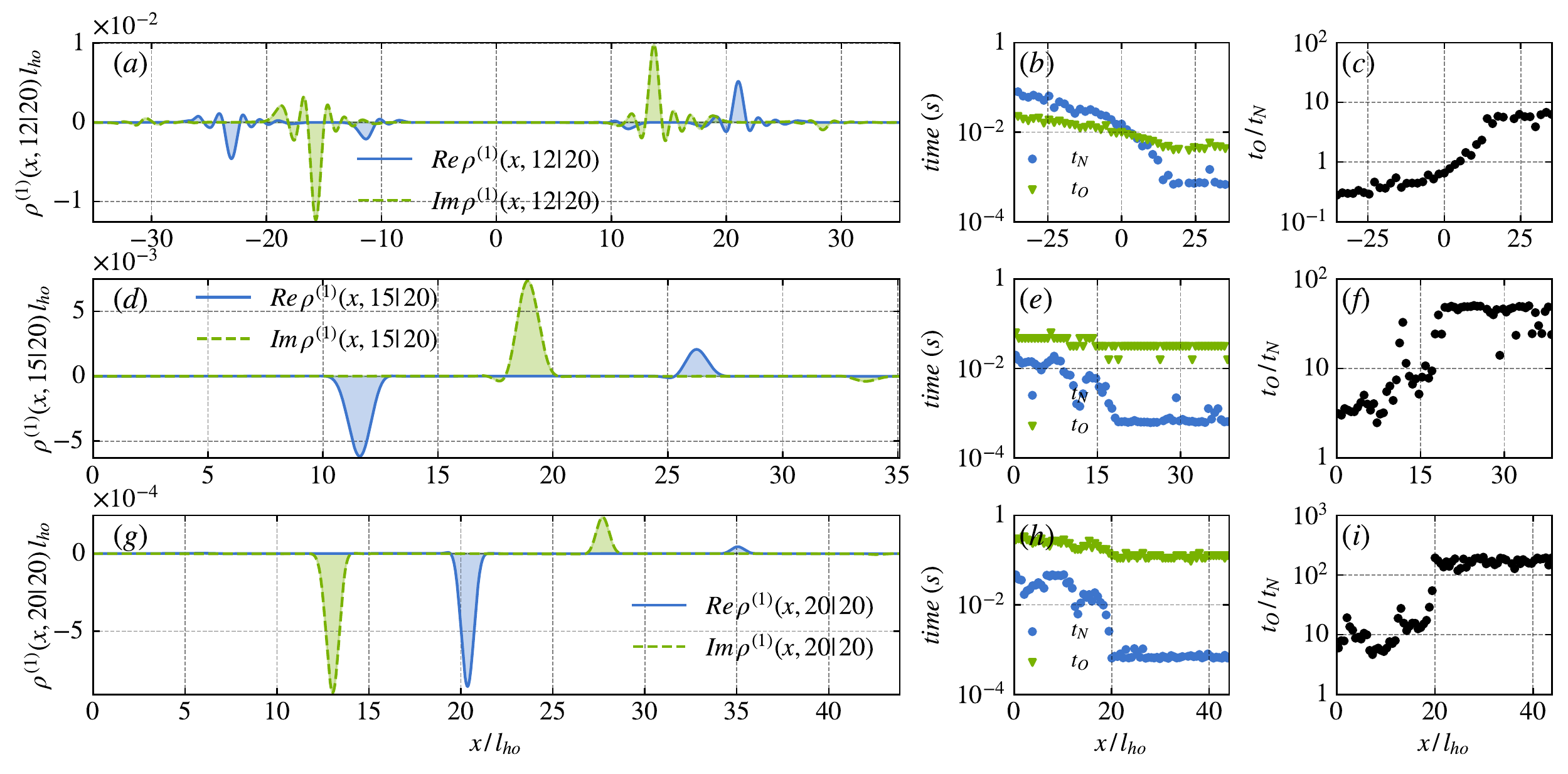}
\caption{Same quantities as in Fig.~\ref{figt2} for $t=20\Delta t$ and $y=12$ (a), $y=15$ (d) and $y=20$ (g). }
\label{figt20}
\end{figure*}

In this Appendix we compare the  efficiencies of the two numerical methods presented in Sect.~\ref{s42} and Sect.~\ref{s43}
which can be used to compute the reduced density matrices of impenetrable anyons. We consider the case when the overlaps of
the wave-functions cannot be calculated analytically as in the  QNC setup (see Sect.~\ref{s5}).

Both methods require the knowledge of the SP wave-functions which can be calculated either analytically (in some cases) or by numerically
solving the time-dependent Schr\"{o}dinger equation. In general the wave-functions are highly oscillatory functions and in order to  accurately calculate the RDM
they need to be computed on a fine grid in an appropriate chosen domain. If the wave-functions need to be evaluated for values different from the gridpoints
we will use interpolation. If we have computed such a sampling of the wave-functions on the equidistant grid $X=(x_1,\cdots, x_n)$ ($n$ is the number
of gridpoints) then we can very easily obtain a sampling of the free fermionic RDM on $X\times X$ which then can be  used for
the 2D interpolation of $\rho_{F}^{(1)}(x,y|\, t)$. In order to see this consider the matrix with elements $A_{ij}=f_i^{1/2}\phi_i(x_j,t)$ where
$f_i$ is the Fermi-Dirac function and $i=0,\cdots, M-1\, , \ j=1,\cdots, n$ (this is the $M\times n$ matrix constructed by putting on row $i$ the wave-function $\phi_i$
evaluated on the grid $X$ and multiplied by $f_i^{1/2}$). Then the $n\times n$ values of $\rho_F^{(1)}(x,y|\,t)$ for $x,y\in X\times X$ are given
by the matrix $A^T A^*$ where $A^T$ is the transposed matrix and $A^*$ is the complex conjugate. We point out that the computation of the free fermionic RDM
 on the 2D grid from the sampled wave-functions can be done very quickly (even in the case of the largest grid that we have employed with $n=3600$
the construction of the matrices and the matrix product took less than a second on a normal laptop).

A correct evaluation of the efficiency of the two methods should take into account, in addition to the floating-point operations required
in each case, the number of calls required by each method for $\phi_i(x,t)$ or $\rho_F^{(1)}(x,y|\,t)$ when $x\ne X$  or $x,y\ne X\times X$.
This is because while function interpolation is very efficient it is still time consuming compared with a floating-point operation. On our system
the  time for a call of an interpolated wave-function was $t_\phi\sim [1.5, 2.5]\times 10^{-6} s$ and for an interpolated fermionic
RDM $t_\rho\sim [4,5]\times 10^{-6} s$.

\textit{Analysis of the method based on the overlaps}. In general the most time consuming component of this method is the computation of the wave-functions
overlaps. Needless to say, the utilization of a general purpose adaptive subroutine, while very accurate, would be time consuming. In order to minimize
the number of calls to the interpolated wave-functions it is preferable to use a quadrature rule whose points and weights can be computed accurately and fast.
We have used the Clenshaw-Curtis quadrature which has a computational cost of  $O(p\, \log\, p)$ using Waldvogel's FFT algorithm \cite{Wald} ($p$ is the number of
points of the quadrature). It can be argued that a more suitable choice of quadrature would be Gauss-Legendre but for smooth functions it seems that
Clenshaw-Curtis performs as well \cite{Tref} and our numerical experiments confirm this hypothesis. Using a quadrature with $p$ points in order to compute the
overlaps we need $M p$ calls of the interpolated wave-functions and then $M(M+1) p/2$ multiplications (only $M(M+1)/2$ integrals are independent the
rest can be obtained from complex conjugation). The computation of the inverse and the determinant of $\boldsymbol{P}$ both require $O(M^3)$ operations,
$\boldsymbol{Q}$ requires $M^2$ multiplications while the summation in (\ref{rhot}) require $2 M$ calls of the interpolated wave-functions and $M^2$ multiplications.
Therefore the evaluation time of the
the truncated basis method for a quadrature with $p$ points and truncation level $M$ is approximately
\begin{align}\label{timeT}
t_O=&\left(p\log p+2 M^3 +  2 M^2 +\frac{M(M+1)}{2}p\right)t_F\nonumber\\
&\qquad\qquad\qquad\qquad\qquad +M(p+2)t_\phi\, ,
\end{align}
where $t_F$ is the time required by a floating-point operation.

\textit{Analysis of Nystr\"{o}m's method}. If we use a quadrature with $m$ points due to the fact that $\rho^{(1)}(x,y|\, t)=(\rho^{(1)}(y,x)|\, t))^*$
the construction of the matrix $\rho^{(1)}(\lambda_i,\lambda_j|\, t)$ requires $m(m+1)/2$ calls of the interpolated RDM and the solution of the linear system (\ref{i21})
require $O(m^3)+O(m^2)$ operations (the $O(m^2)$ comes from the multiplication of the aforementioned matrix with $w_j$). The Fredholm determinant also
requires require $O(m^3)+O(m^2)$ operations and the interpolation formula (\ref{resolventinterp}) requires  $O(m)$ operations  and $m+1$ calls of the interpolated RDM.
Therefore, the approximate evaluation time of Nystr\"{o}m's method is
\begin{align}\label{timeN}
t_N=&(m\log m+ 2m^3+2m^2+m)t_F\nonumber\\
&\qquad\qquad\qquad\qquad +\frac{1}{2}(m+1)(m+2)t_\rho\, .
\end{align}
Taking into account that $t_\rho\sim 2t_\phi$ and that it is sensible to assume that $p\sim m$ (this was true in all our numerical investigations) from
(\ref{timeT}) and (\ref{timeN}) we can already estimate which method is more competitive. Because we are dealing with smooth functions and kernels in
general the number of quadrature points required increases monotonically with $|x-y|$. For  $m<M$ (\ref{timeN}) is smaller than (\ref{timeT}) which shows
that Nystr\"{o}m's method is more efficient for small to moderate values of $|x-y|$ and moderate to large number of particles (or equivalently at higher temperatures).
When $M$ is small and $|x-y|$ is large the overlaps method is more efficient.

The middle column of Fig.~\ref{figt2} presents the evaluation time of $\rho^{(1)}(x,2|\, 2\Delta t)$ at different temperatures for a bosonic system of $N=10$ particles in the
quantum Newton's cradle setup characterized by $l_{HO}=0.56, \omega=3, q=4\pi,$ $A=1.5$ and $\Delta t= \pi/40 \omega$. The evaluation times are obtained by increasing the number of quadrature points
until the relative difference between two successive evaluations is smaller than $0.5\times 10^{-3}$. At zero temperature ($M=N-1$) from Fig.~\ref{figt2} (b) we can see that for
small to moderate values of the spatial separation Nystr\"om's method is more efficient $(t_N<t_O)$ but the situation is reversed for large values of $|x-2|$. The last column of
Fig.~\ref{figt2} presents the results for the $t_O/t_N$. If $t_O/t_N>1$  then Nystr\"{o}m's method is more efficient and when $t_O/t_N<1$ the opposite is true. Already for
$\theta_0=0.011$ when $M=30$ we see from Fig.~\ref{figt2}(e) that $t_O/t_N>1$ for all values of $x$ shown (outside of this interval the RDM is negligible) and for $\theta_0=0.0025$ (Fig.~\ref{figt2}(h))
when $M=60$ Nystr\"om's method is almost everywhere more than ten times faster than the method based on the overlaps. Fig.~\ref{figt20} presents similar results for the same
setup at $t=20\Delta t$ and $y=12$ $(\theta_0=0)$,  $y=15$ $(\theta_0=0.011)$ and $y=20$ $(\theta_0=0.025)$. Even though the support of the   RDM is now larger the same conclusions can be drawn:
with increasing $M$ and temperature Nystr\"{o}m's method is far more efficient than the overlaps method.

\end{document}